\documentclass[fleqn,usenatbib]{mnras}

\usepackage{newtxtext,newtxmath}

\usepackage[T1]{fontenc}
\usepackage{ae,aecompl}

\usepackage{graphicx}   
\usepackage{amsmath}    
\usepackage{amssymb}    

\usepackage{bm, xcolor}
\usepackage[all]{hypcap}
\usepackage{times}

\def\cm3{\hbox{cm$^{-3}$}}
\newcommand\Msun{\,\rmn{M}_{\sun}}
\newcommand\Zsun{\,\rmn{Z}_{\sun}}
\newcommand\Gyr{\,\rmn{Gyr}}
\newcommand\Myr{\,\rmn{Myr}}
\newcommand{\yr}{\,{\rm yr}}

\newcommand\pc{\,\rmn{pc}}
\newcommand\kpc{\,\rmn{kpc}}

\newcommand\Mag{\,\rmn{mag}}
\newcommand\Arcsec{\,\rmn{arcsec}}
\newcommand\Mvir{M_\rmn{vir}}

\newcommand{\muv}{M_{\rm UV}}

\newcommand{\gadget}   {\textsc{gadget3}}
\newcommand{\subfind}  {\textsc{subfind}}

\newcommand{\cmcubed}  {\,{\rm cm}^{-3}}

\newcommand{\cMpc}     {\,{\rm cMpc}}

\title[Predictions for high-$z$ GC UVLFs in E-MOSAICS]
{The evolution of the UV luminosity function of globular clusters in the E-MOSAICS simulations}
\author[Pfeffer et al.] {Joel Pfeffer,$^{1}$ Nate Bastian,$^{1}$ Robert A. Crain,$^{1}$  J.~M.~Diederik Kruijssen,$^{2}$ \newauthor Meghan E. Hughes,$^{1}$ and Marta Reina-Campos$^{2}$\\
$^{1}$Astrophysics Research Institute, Liverpool John Moores University, 146 Brownlow Hill, Liverpool L3 5RF, UK\\
$^{2}$Astronomisches Rechen-Institut, Zentrum f\"ur Astronomie der Universit\"at Heidelberg, M\"onchhofstra{\ss}e 12-14, 69120 Heidelberg, Germany.\\
}
\date{Accepted 2019 June 4. Received 2019 May 17; in original form 2018 October 22}
\pagerange{\pageref{firstpage}--\pageref{lastpage}}
\pubyear{2019}
\begin{document}
\maketitle
\label{firstpage}
\begin{abstract}
We present the evolution of the rest-frame ultraviolet (UV) properties of the globular cluster (GC) populations and their host galaxies formed in the E-MOSAICS suite of cosmological hydrodynamical simulations. 
We compute the luminosities of all clusters associated with 25 simulated Milky Way-mass galaxies, discussed in previous works, in the rest-frame UV and optical bands by combining instantaneous cluster properties (age, mass, metallicity) with simple stellar population models, from redshifts $z=0$ to $10$.  
Due to the rapid fading of young stellar populations in the UV, most of the simulated galaxies do not host GCs bright enough to be individually identified in deep Hubble Space Telescope (HST) observations, even in highly magnified systems.  
The median age of the most UV-luminous GCs is $<10$~Myr (assuming no extinction), increasing to $\gtrsim100$~Myr for red optical filters.  
We estimate that these GCs typically only contribute a few per cent of the total UV luminosity of their host galaxies at any epoch. 
We predict that the number density of UV-bright proto-GCs (or cluster clumps) will peak between redshifts $z=1-3$. 
In the main progenitors of Milky Way-mass galaxies, $10$-$20$ per cent of the galaxies at redshifts $1 \lesssim z \lesssim 3$ have clusters brighter than $\muv < -15$, and less than $10$ per cent at other epochs.
The brightest cluster in the galaxy sample at $z>2$ is typically $\muv \sim -16$, consistent with the luminosities of observed compact, high-redshift sources.
\end{abstract}
\begin{keywords}
stars: formation -- globular clusters: general -- galaxies: formation -- galaxies: evolution -- galaxies: star clusters:   general -- methods: numerical
\end{keywords}

\section{Introduction}
\label{sec:intro}

Combining the resolving power of the Hubble Space Telescope (HST) with the magnification of gravitational lensing, it is now possible to peer into galaxies during the epoch of globular cluster (GC) formation. 
Initial studies with HST of lensed objects (HST Frontier Fields, \citealt{Lotz_et_al_17}; SGAS-HST, Gladders et al. in prep.) have revealed a small population of sources with properties consistent with that of young GCs and star cluster complexes ($R_{\rm eff} < 50$~pc and $M > 10^5 \Msun$) \citep{Kawamata_et_al_15, Kawamata_et_al_18, Livermore_et_al_15, Bouwens_et_al_17c, Elmegreen_and_Elmegreen_17, Hernan-Caballero_et_al_17, Johnson_T_et_al_17a, Johnson_T_et_al_17b, Vanzella_et_al_17a, Vanzella_et_al_17b, Vanzella_et_al_19}. 
These objects include both apparently isolated sources and sub-clumps within larger systems.
While possible analogues of young GCs ($R_{\rm eff}\sim3-10$~pc, $M\sim10^5 - 10^8 \Msun$; i.e. ``young massive clusters'') have been found and studied in detail in the local Universe \citep[e.g.][and references therein]{Holtzman_et_al_92, Portegies-Zwart_McMillan_and_Gieles_10, Kruijssen_14, Longmore_et_al_14, Adamo_and_Bastian_18}, the opportunity to study directly the formation of the (now ancient) GCs is particularly exciting, because it enables direct tests of theories for the formation and co-evolution of galaxies and their GC populations.

The initial masses (luminosities) of young GCs are of interest as some theories for the origin of ``multiple stellar populations'' within GCs invoke extreme cluster mass loss, requiring in these scenarios that GCs were factors of 10-100 times more massive at birth than at present \citep[e.g.][]{D_Ercole_et_al_08,Krause_et_al_13}.  While such self-enrichment scenarios are in conflict with a number of other constraints (see \citealt{Bastian_and_Lardo_18} for a recent review) they have remained popular owing to a lack of compelling alternatives.  Clearly, if GCs were significantly more massive at birth than at present, then they would be much brighter and potentially much easier to detect in the high-redshift Universe.

Of particular interest is the evolution of the GC rest-frame UV luminosity function (LF), as it potentially enables a strong test of GC formation theories (e.g. \citealt{Katz_and_Ricotti_13, Bouwens_et_al_17c, Boylan-Kolchin_17, Renzini_17}; see \citealt{Forbes_et_al_18} for a recent review on GC formation).
It informs whether GCs, especially the metal-poor sub-population, only form at high redshift ($z > 6$), or exhibit a formation history more reflective of the star formation in their host galaxy, with a broader redshift distribution.  
This in turn has a strong bearing on whether GCs may have played a significant role in reionization \citep[e.g.][]{Ricotti_02, Griffen_et_al_13, Katz_and_Ricotti_13, Katz_and_Ricotti_14, Boylan-Kolchin_17, Boylan-Kolchin_18}.

An additional consideration when studying young GCs is that, at least in the local Universe, massive clusters seldom form in isolation, but more commonly as part of larger stellar/cluster complexes \citep[e.g.][]{Efremov_and_Elmegreen_98, Zhang_Fall_and_Whitmore_01}.  
These complexes have half-light radii of 10s to 100s of pc, compared to $~\sim1-10$~pc for individual clusters, meaning that unless the resolution of the imaging is better than $10-20$~pc, observations of young clusters can be subject to significant contamination from nearby young clusters and field stars.
Images of high-redshift galaxies also often show clumpy distributions in the young stars (and ionized gas), \citep[e.g.][]{Elmegreen_et_al_05, Elmegreen_et_al_07, Elmegreen_et_al_09, Shapiro_Genzel_and_Forster_Schreiber_10, Forster_Schreiber_et_al_11, Genzel_et_al_11, Guo_et_al_12, Guo_et_al_15, Adamo_et_al_13, Shibuya_et_al_16}, though the most massive ``clumps'' may have overestimated masses due to blending of smaller clumps at the resolution limits \citep{Dessauges-Zavadsky_et_al_17, Rigby_et_al_17, Cava_et_al_18}.
The mass function of the clumps is consistent with a power law of slope $\approx -2$ \citep{Dessauges-Zavadsky_and_Adamo_18}, suggesting they are formed by fragmentation in turbulent, hierarchical gaseous discs \citep{Elmegreen_and_Falgarone_96} similar to molecular clouds in local galaxies \citep[e.g.][]{Stutzki_et_al_98, Dickey_et_al_01, Freeman_et_al_17}. 
Therefore, this suggests that cluster complexes may also be common in high-redshift galaxies.
If similar processes are at play in both the low- and high-redshift Universe, then cluster complexes might artificially inflate estimates of young GC masses at high redshifts.

A number of recent studies have interpreted high-redshift rest-frame UV observations in the context of the present day GC populations of the Milky Way and nearby galaxies.  
Assuming metal-poor GCs form between redshifts of $z=10$ and $z=4$, \citet{Boylan-Kolchin_18} concluded that models that assume large mass loss factors for GCs ($>10$ times their current mass) may already exceed the observed high-redshift UV LFs, implying GCs cannot have been significantly more massive at birth \citep[see also][]{Boylan-Kolchin_17, Bouwens_et_al_17c}.
Based on reconstructing the evolution of the Fornax dwarf spheroidal galaxy from its present day properties, \citet{Zick_Weisz_and_Boylan-Kolchin_18} proposed that young GCs at high redshift may be significantly brighter (10-100 times the flux) in the UV than their host galaxy, depending on formation time of GCs relative to the galaxy star formation rate (SFR). 
If true, high-redshift observations in the rest-frame UV might preferentially detect young GCs, rather than the host galaxy.

Semi-analytic models placing GC formation into the context of cosmological, hierarchical galaxy assembly have had various successes in explaining the properties of GC populations, such as metallicity distributions, specific frequencies and the ``blue tilt'' \citep{Beasley_et_al_02, Bekki_et_al_08, Prieto_and_Gnedin_08, Griffen_et_al_10, Muratov_and_Gnedin_10, Tonini_13, Katz_and_Ricotti_14, Li_and_Gnedin_14, Kruijssen_15, Choksi_Gnedin_and_Li_18}.
However, in general, such models lack detailed information concerning the baryonic processes in galaxies necessary for contrasting star cluster and field star populations in high-redshift galaxies. 
Recently, cosmological hydrodynamical simulations of galaxy formation have begun to incorporate models of GC formation, either through subgrid treatments or by directly resolving cluster formation \citep{Kravtsov_and_Gnedin_05, Ricotti_Parry_and_Gnedin_16, Li_et_al_17, Kim_et_al_18, P18}, making such comparisons possible.

In this work, we investigate the properties of young GCs at high redshift in the context of the E-MOSAICS simulations.
The E-MOSAICS project \citep[MOdelling Star cluster population Assembly In Cosmological Simulations within EAGLE,][]{P18, K19} is a suite of simulations that include star cluster formation, evolution and disruption, dependent on the local conditions in which they form and evolve, within the EAGLE (Evolution and Assembly of GaLaxies and their Environments) hydrodynamical simulations of galaxy formation \citep{S15, C15}.  
The simulations adopt a model for GC formation, based on models for young star cluster formation, that has been widely tested against observations of massive cluster formation in nearby galaxies \citep{Kruijssen_12, Adamo_et_al_15, Johnson_et_al_16, Reina-Campos_and_Kruijssen_17, Messa_et_al_18_II, Pfeffer_et_al_19}. 
In this model, GCs are the remnants of normal star cluster formation at high redshift.
The model has been extensively tested and benchmarked \citep{P18} and successfully applied to the Milky Way to interpret the age-metallicity relations of its GC population and reconstruct its formation and assembly history \citep{K19, K19b}, as well as to reproduce the metallicities and ages of star clusters in Milky Way satellite galaxies \citep{Hughes_et_al_19} and the ``blue tilt'' of GC populations \citep{Usher_et_al_18}.
In the present paper, we use the same set of $25$ zoom simulations of Milky Way-mass galaxies and their satellite populations presented by \citet{K19} to study the rest-frame UV properties of the young GCs at high redshift.  
These are used to interpret observations of lensed galaxies and their cluster/complex populations as well as to make predictions for future observations.

This paper is organised as follows: 
In Section \ref{sec:sims} we briefly describe the E-MOSAICS simulations and the method for calculating the luminosities of clusters in different filters.
Section \ref{sec:lf} presents the main results of this work on the cluster UV magnitudes, luminosity functions and fraction of UV flux in a galaxy contributed by clusters.
Finally, we end with the discussion and conclusions in Sections \ref{sec:discussion} and \ref{sec:conclusions}, respectively.

\section{Simulations}
\label{sec:sims}

\subsection{E-MOSAICS}

The E-MOSAICS simulations are described in detail by \citet{P18} and \citet{K19}, and we refer the reader to those works for a comprehensive overview of the models.  Here we present a brief overview of the simulations, focussing only on the elements germane to the present study. 

The E-MOSAICS project is a suite of hydrodynamical simulations of galaxy formation in the $\Lambda$ cold dark matter cosmogony that couple the MOSAICS subgrid model for star cluster formation and evolution \citep{Kruijssen_et_al_11,P18} to the EAGLE galaxy formation model \citep{S15,C15}.
The EAGLE model includes routines describing the subgrid physics for radiative cooling \citep{Wiersma_Schaye_and_Smith_09}, star formation \citep{Schaye_and_Dalla_Vecchia_08}, stellar mass loss \citep{Wiersma_et_al_09}, energy feedback from star formation \citep{Dalla_Vecchia_and_Schaye_12}, gas accretion on to and mergers of supermassive black holes \citep{Rosas_Guevara_et_al_15} and active galactic nuclei feedback \citep{Booth_and_Schaye_09}. 
In EAGLE, the subgrid efficiencies of the stellar and black hole feedback are calibrated to reproduce the present-day galaxy stellar mass function, galaxy size-mass relation and black hole masses.
The simulations are run with a significantly modified version of the $N$-body TreePM smoothed particle hydrodynamics (SPH) code \gadget\ \citep[last described by][]{Springel_05}.

The MOSAICS star cluster formation and evolution model couples to the EAGLE model in a subgrid manner, such that a population of star clusters forms a subgrid component of each newly-formed star particle, with the clusters inheriting the properties of their host particle (e.g. ages, metallicities, positions).
The clusters do not affect the evolution of the simulations, thus avoiding any recalibration of the EAGLE model.
Both the initial properties of clusters and their subsequent evolution are governed by the local conditions of the host particle, such as the local ambient gas and dynamical properties.
The masses of the newly-formed clusters are decoupled from the mass of the stellar particle, such that cluster masses are not dependent on the simulation resolution and a stellar particle may host clusters with masses larger than itself.

MOSAICS adopts a cluster formation model that reproduces the properties of young star cluster populations in nearby galaxies \citep{Kruijssen_12, Reina-Campos_and_Kruijssen_17, Pfeffer_et_al_19}.
In the model, star clusters form with a \citet{Schechter_76} initial cluster mass function with a power law slope of $-2$.
Clusters are sampled from the mass function between masses of $10^2$ and $10^8 \Msun$, though only clusters more massive than $5 \times 10^3 \Msun$ are evolved to reduce memory requirements.
The initial properties of the star cluster population of each particle are then determined by two main properties: the cluster formation efficiency \citep[CFE or $\Gamma$, i.e. the fraction of star formation in bound star clusters][]{Bastian_08} and the exponential truncation mass of the \citet{Schechter_76} initial cluster mass function, both of which vary with the local environment.
The CFE is determined from the \citet{Kruijssen_12} model and varies as a function of the local natal gas properties (namely the gas pressure in the E-MOSAICS formulation).
The exponential truncation mass is determined from a local formulation of the \citet{Reina-Campos_and_Kruijssen_17} model, where the truncation mass generally increases with gas pressure but decreases in regions with high centrifugal forces (i.e. near the centres of galaxies).

The E-MOSAICS model does not make any assumptions about cluster mass loss in order to satisfy models for the origin of multiple populations.
In the terminology of \citet{Boylan-Kolchin_17}, this means that we do not adopt a constant $\xi$ (defined as the ratio of the mass at formation to the present day mass of a GC), but instead model the evolution of this quantity self-consistently due to mass loss by stellar evolution, two-body relaxation and tidal shocks with the surrounding environment. 
For old clusters ($>10 \Gyr$) at $z=0$, $\xi$ is largely a function of cluster mass, with massive clusters ($M>10^5~\Msun$) having $\xi \approx 1.8$ simply due to stellar evolutionary mass loss \citep[see][]{Reina-Campos_et_al_18a}.
Large ($>10$) values of $\xi$\ have been shown to be incompatible with observations of the GC populations of a number of nearby dwarf galaxies \citep[Fornax dwarf spheroidal, WLM, IKN,][]{Larsen_Strader_and_Brodie_12, Larsen_et_al_14}, as well as the Milky Way GC population \citep[see][for a review]{Bastian_and_Lardo_18}.  
Additionally, they are incompatible with the observed low-mass end of the stellar mass function of present day GCs, which is sensitive to mass loss \citep[e.g.][]{Kruijssen_09, Webb_and_Leigh_15}.
We investigate such mass-loss scenarios in the context of E-MOSAICS in \citet{Reina-Campos_et_al_18a}.
Finally, the removal of star clusters by dynamical friction in their host galaxy is treated in post-processing and applied at every snapshot \citep{P18}.

In this work, we use the volume-limited sample of 25 simulated galaxies with Milky Way-mass haloes ($\Mvir \approx 10^{12} \Msun$), which were drawn from the high resolution $25 \cMpc$ volume EAGLE simulation \citep[Recal-L025N0752,][]{S15} and resimulated in a zoom-in fashion with the E-MOSAICS model \citep[see][]{P18, K19}.
The simulations were rerun with the same parameters as the parent volume, using a \citet{Planck_2014_paperI} cosmology, the `recalibrated' EAGLE model \citep[see][]{S15} and initial baryonic particle masses of $\approx 2.25 \times 10^5 \Msun$.
For each simulation, 29 snapshots were output between redshifts of $z=20$ and $z=0$.
Bound galaxies (subhaloes) at each snapshot were determined using the \subfind\ algorithm \citep{Springel_et_al_01, Dolag_et_al_09} and subhalo merger trees were created using the method described in \citet{P18}.

\subsection{Cluster luminosities} \label{sec:SSPs}

At each epoch of `observation' we use the present mass, and hence luminosities, of all GCs in the galaxies.
We estimate each cluster's luminosity in six rest-frame passbands, $M_{\rm 1500}$ (referred to as $\muv$) and the five SDSS filters $M_{\rm u}$, $M_{\rm g}$, $M_{\rm r}$, $M_{\rm i}$, and $M_{\rm z}$.  
To do this, we use the clusters' current (at any given snapshot) age, metallicity and mass in combination with predictions from the \textsc{fsps} model \citep*{Conroy_Gunn_and_White_09,Conroy_and_Gunn_10} using the Miles spectral library \citep{Sanchez-Blazquez_et_al_06} and Padova isochrones \citep{Girardi_et_al_00,Marigo_and_Girardi_07,Marigo_et_al_08}. 
We use the default \textsc{fsps} parameters, assume simple stellar populations for clusters and adopt a \citet{Chabrier_03} stellar initial mass function (consistent with the simulations).
For each filter, mass-to-light ratios for the clusters were calculated by linearly interpolating from the grid in ages and total metallicities $\log(Z/\Zsun)$.
All magnitudes are in the AB system.  

In order to estimate the effect of extinction when converting our model cluster properties to observations, we adopt the same methodology as \citet{Boylan-Kolchin_18}, namely by adopting two (nearly) limiting cases.  The first is to assume no extinction at all (i.e. that clusters are visible immediately after their formation), the second is to assume that they are fully embedded within an optically thick cloud until a specific age, which we adopt to be $10$~Myr \citep[e.g.][]{Charlot_and_Fall_00}.  This could be due to, for example, SNe Ib,c and IIs clearing the gas from the progenitor GMC on this timescale.  As will be shown, due to the  rapid time evolution of the UV flux of clusters, the age at which clusters become visible (in the UV) strongly influences the resulting luminosity functions.

In order to track how observed properties change as the host galaxy and its GC population form and evolve, we perform analysis of the cluster populations in all 29 snapshots for the simulations, but mainly investigate snapshots at redshifts $z=6, 5, 4, 3, 2, 1, {\rm and}~0.5$ for brevity.
We primarily focus on clusters associated with the main progenitors of each galaxy (i.e. particles bound to the halo according to \subfind\ and excluding bound satellites). 
However, in Section \ref{sec:uv_frac} we briefly investigate the clusters of all progenitors of the galaxies and their $z=0$ satellites.

\section{Population properties}
\label{sec:lf}

\subsection{Luminosity vs. age}

\begin{figure*}
\centering
\includegraphics[width=15cm]{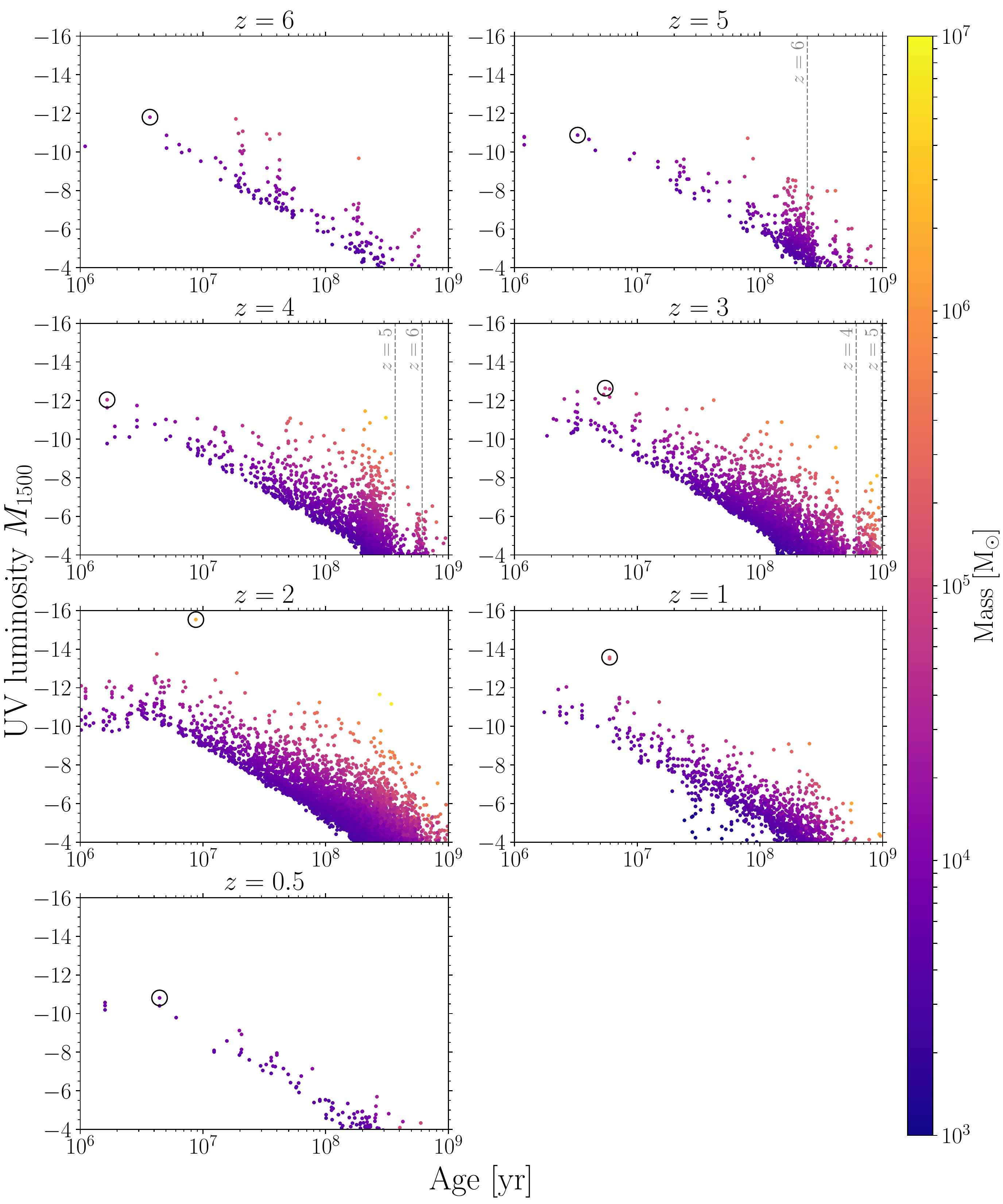}
\caption{The age-$\muv$ plane of the clusters at seven redshifts, labelled at the top of each panel, for galaxy MW00 in the E-MOSAICS suite of simulations. Dashed vertical lines indicate the redshifts for previous panels (i.e. clusters older than this age may be repeated in multiple panels). For the results shown here, no extinction is included.  Only clusters with masses greater than $5\times10^3 \Msun$ (at the epoch of observation) are included and the colour-bar shows the cluster mass.  The brightest cluster in the $\muv$ is circled in each panel, highlighting the fact that in all cases, the most luminous cluster is not the most massive.  The lack of points in the lower-left of each panel is caused by the applied cut in cluster mass, with the slope of the distribution controlled by the fading curve of the stellar models in that filter.  Note that in most cases the brightest cluster has an age $<10$~Myr.}
\label{fig:age_uv}
\end{figure*}

In Fig.~\ref{fig:age_uv} we show the age-$\muv$ plane of the cluster population of the E-MOSAICS galaxy MW00 at seven redshifts.  We circle the brightest cluster (in the UV) in each panel.  The clusters are colour-coded by mass, and the dearth of clusters in the lower-left of each panel is due to the lower cluster mass limit applied to the population ($5\times10^3 \Msun$), which we apply to limit the memory footprint of the simulations.  At magnitudes fainter than $\muv= -10$, the luminosity functions are therefore a combination of the intrinsic LF and incompleteness, hence we restrict our analysis to clusters brighter than this threshold.

A striking feature of Fig.~\ref{fig:age_uv} is the fact that the brightest cluster in the UV is seldom the most massive, with typical masses below $\sim10^5 \Msun$.  
This is caused by the very rapid fading of stellar populations in the UV, driven by the short lifetime of massive stars \citep[e.g. see fig. 2 in][]{Madau_and_Dickinson_14}. 
This result is not unexpected, given the low probability of observing very young (ages of a few tens of megayears) massive clusters within a population described by a power-law or \citet{Schechter_76} mass function \citep[e.g.][]{Gieles_et_al_06b}.
With sufficiently deep imaging, most often observations in the UV would instead primarily observe young, relatively low-mass clusters.
Given a typical detection limit of HST for lensed galaxies of $\muv \sim -14$ \citep{Bouwens_et_al_17b}, only a single GC within MW00 shown in Fig.~\ref{fig:age_uv} would have been detected (at $z=2$ in this case).  
Most of the massive clusters would have been detected if they were observed at precisely the right time ($\muv = -14$ corresponds to a mass detection limit of $\approx 2 \times 10^5 \Msun$ at an age of $1 \Myr$), but it is unlikely to catch them when they are young enough to still be UV bright.
Within MW00, the oldest clusters formed at $z\approx12$ (i.e. $5.7 \times 10^8 \yr$ at $z=6$, top left panel in Fig.~\ref{fig:age_uv}). However, such clusters generally have low masses (initial mass $<10^5 \Msun$) and do not survive for a Hubble time \citep[also see][]{Kruijssen_19}. The first cluster with $M>10^5 \Msun$ forms in the galaxy at $z=7.2$ ($2 \times 10^8 \yr$ at $z=6$).

\begin{figure}
\centering
\includegraphics[width=84mm]{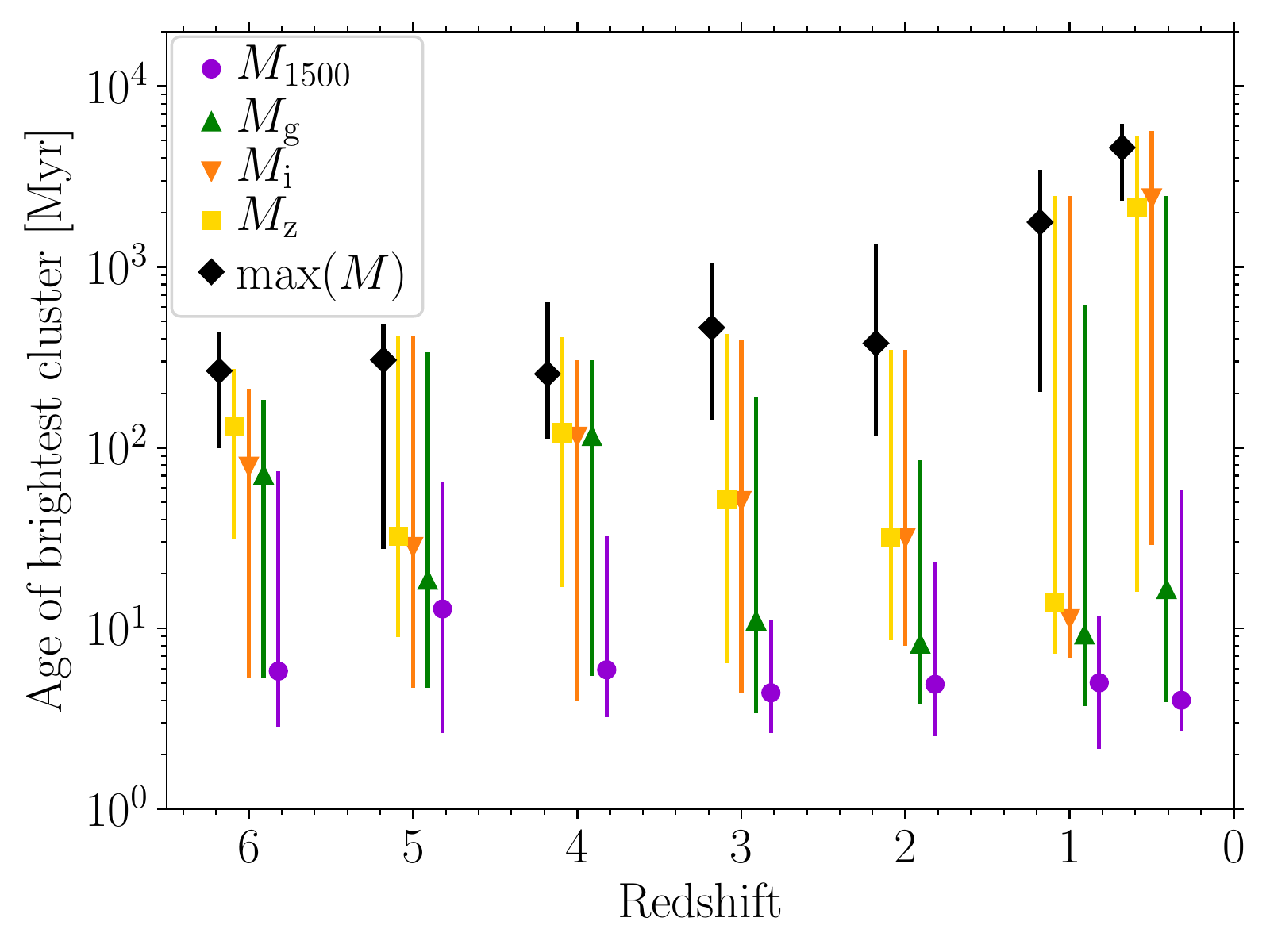}
\caption{The median age of the most massive (black diamonds) and the brightest cluster at each snapshot epoch in the $UV$, $g$, $i$ and $z$ bands, in the absence of extinction, for the 25 Milky Way-mass haloes.  The lines denote the range of the $16^{\rm th}$ and $84^{\rm th}$ percentiles.  If extinction is included, the median age of the $\muv$ increases, as well as the $g$-band, but the $i$ and $z$ are negligibly affected.  Hence, observations of cluster populations in the rest-frame UV provide a highly biased sample, containing essentially only clusters formed in the past $\sim10$~Myr.  Observations in the rest-frame red-optical or near-infrared give a more representative view of the population, except near the peak of cluster formation in the galaxies ($z \sim 1$-$2$), where bright clusters are dominated by young clusters, even in the near-infrared.}
\label{fig:median_age}
\end{figure}

We quantify this further in Fig.~\ref{fig:median_age} where we show the median age of {the most massive cluster and} the brightest cluster in the rest-frame $UV$, $g$, $i$ and $z$-bands.  In the case of no extinction, the average age of the brightest UV cluster is $<10$~Myr, but this varies strongly with wavelength.  In the $z$-band, for example, the median age of the brightest cluster is closer to $100$~Myr, meaning that a cluster sample identified in redder bands is less biased towards very recent star-formation.  In such cases, there is a greater chance that the brightest cluster is among the most massive clusters.
This is not the case near the peak of cluster formation in the galaxies ($z \sim 1$-$2$). 
At these epochs, the brightest clusters are dominated by young clusters ($\sim 10 \Myr$), even in the near-infrared.
However at early ($z\gtrsim3$) and late ($z<1$) times, redder bands are more representative of the most massive clusters.

\subsection{Luminosity functions}

\subsubsection{Without extinction}

\begin{figure*}
\centering
\includegraphics[width=6cm]{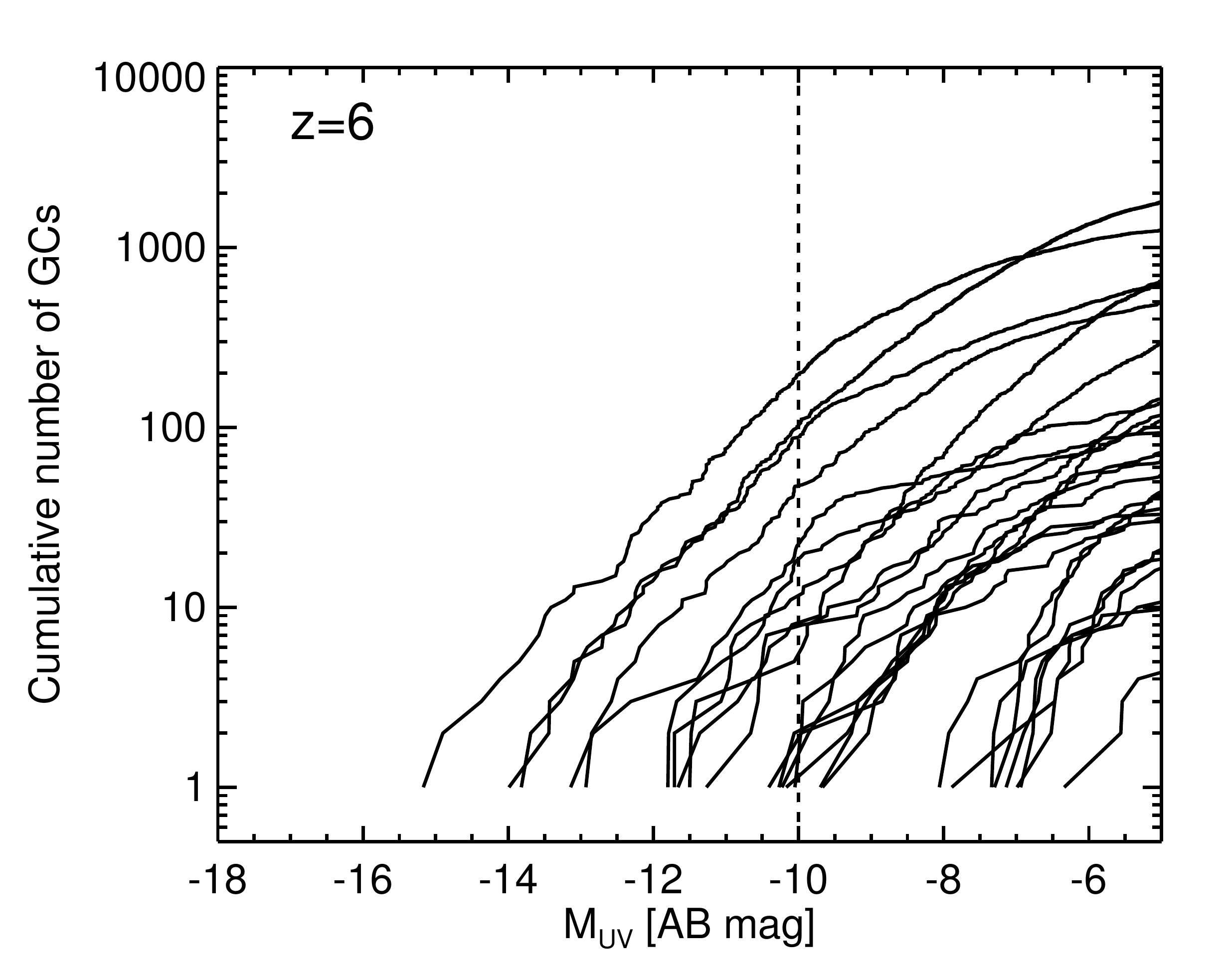}
\includegraphics[width=6cm]{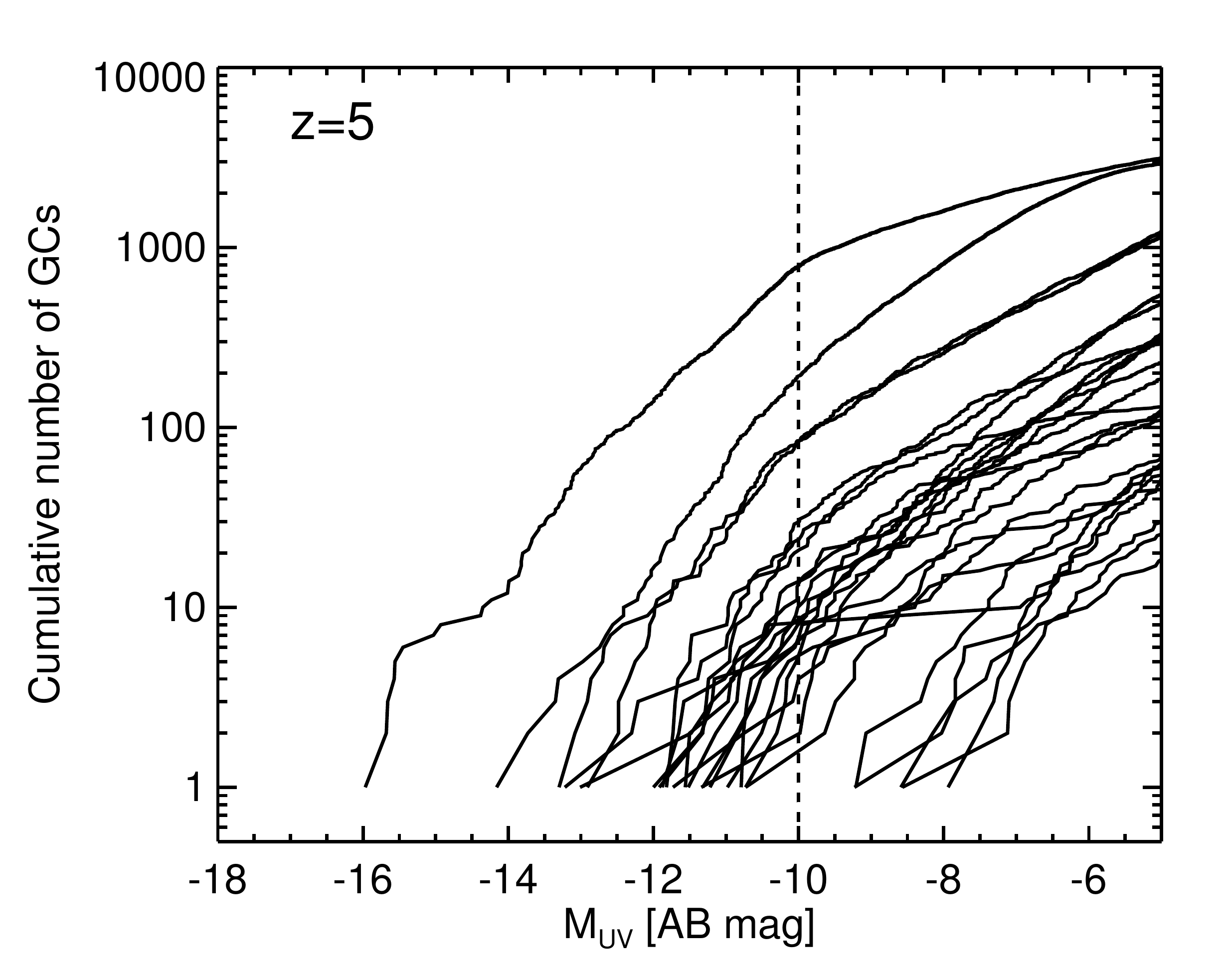}
\includegraphics[width=6cm]{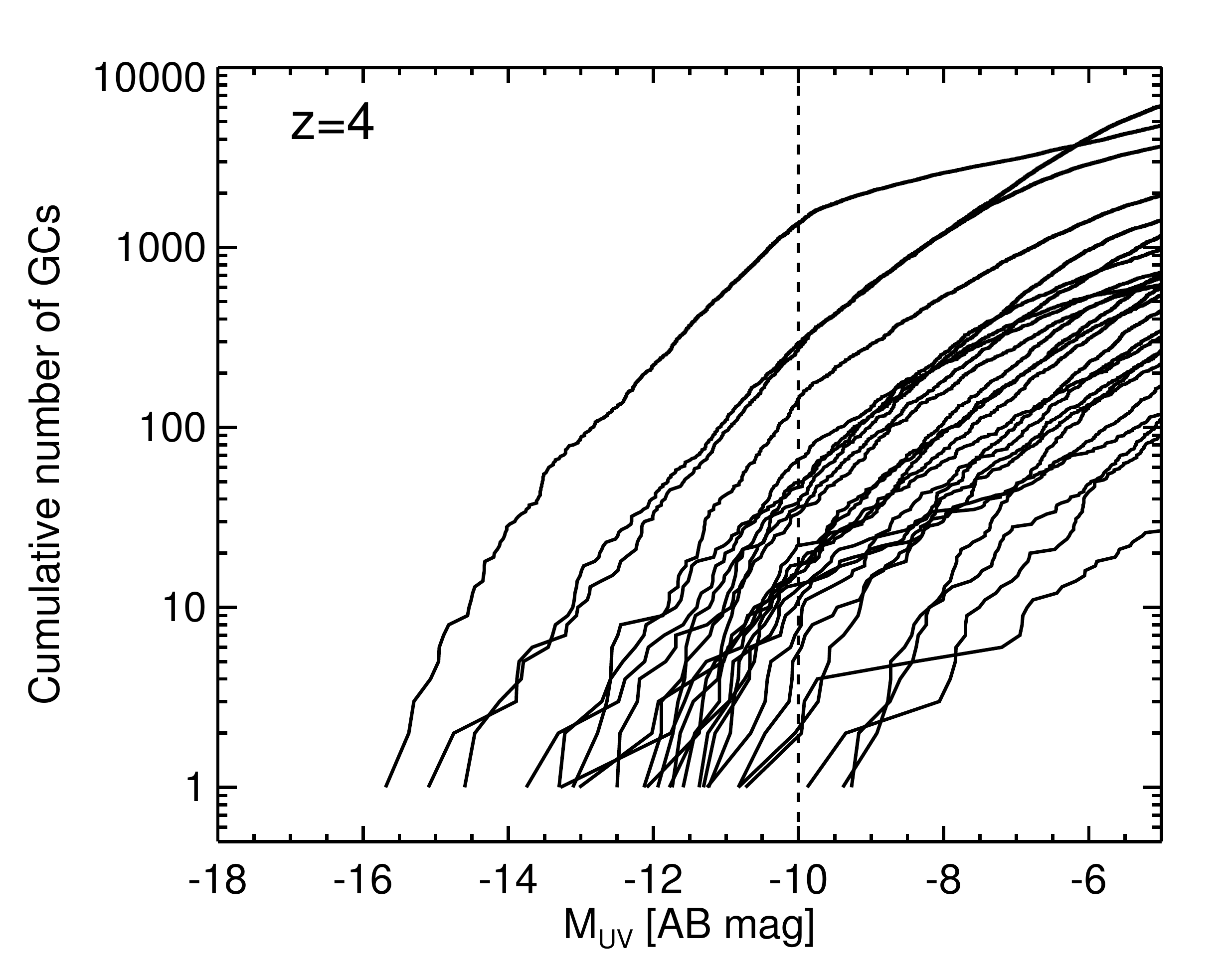}
\includegraphics[width=6cm]{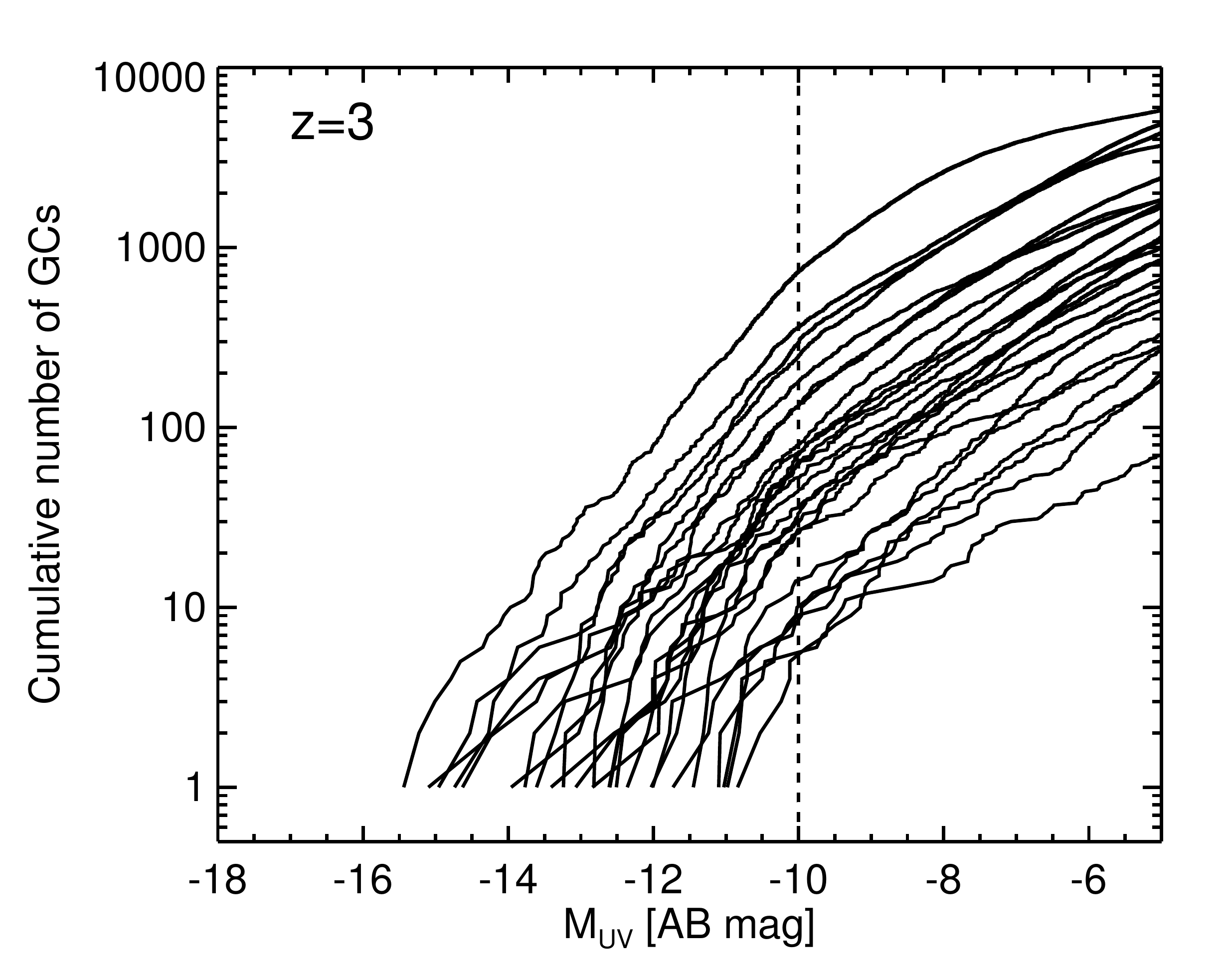}
\includegraphics[width=6cm]{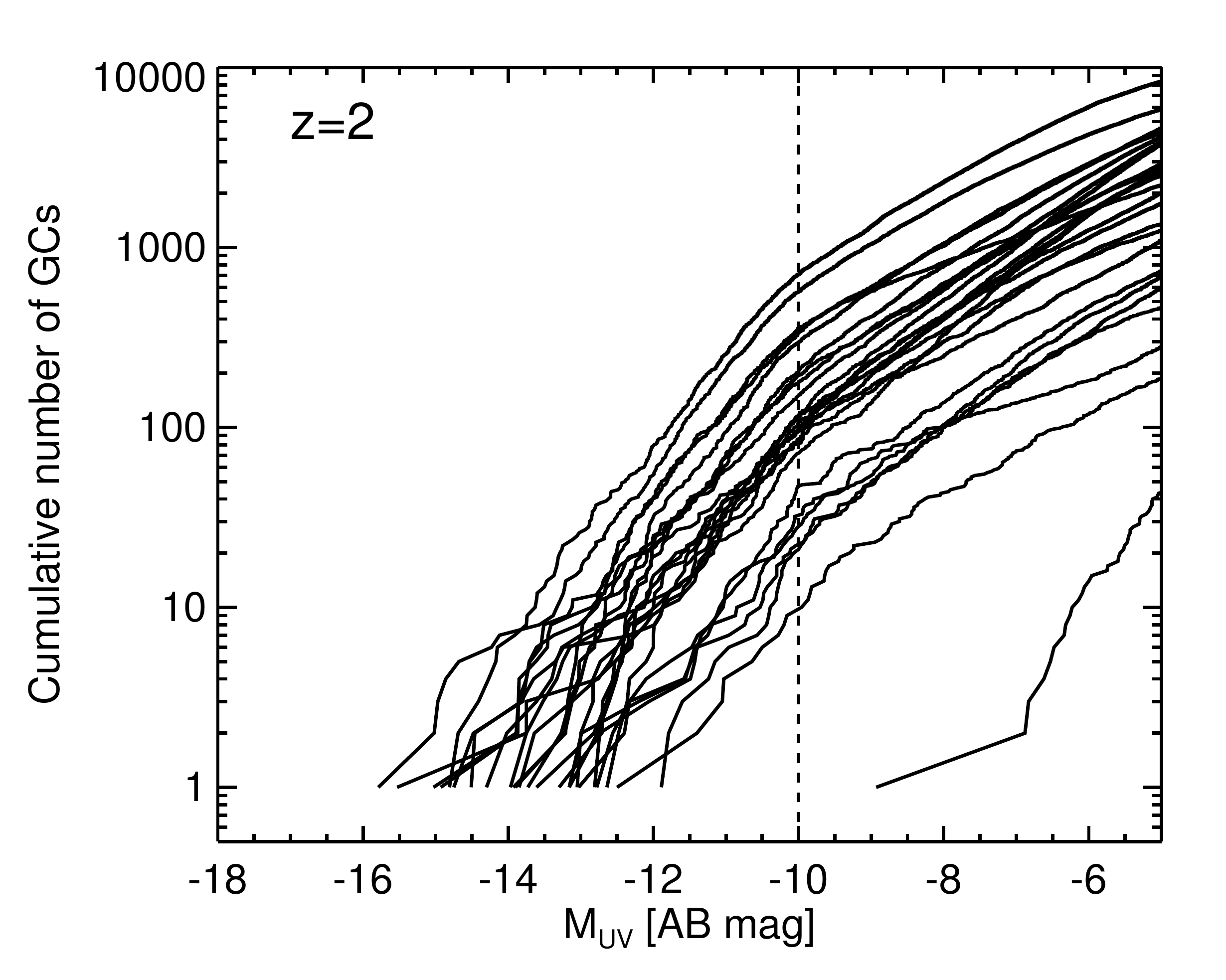}
\includegraphics[width=6cm]{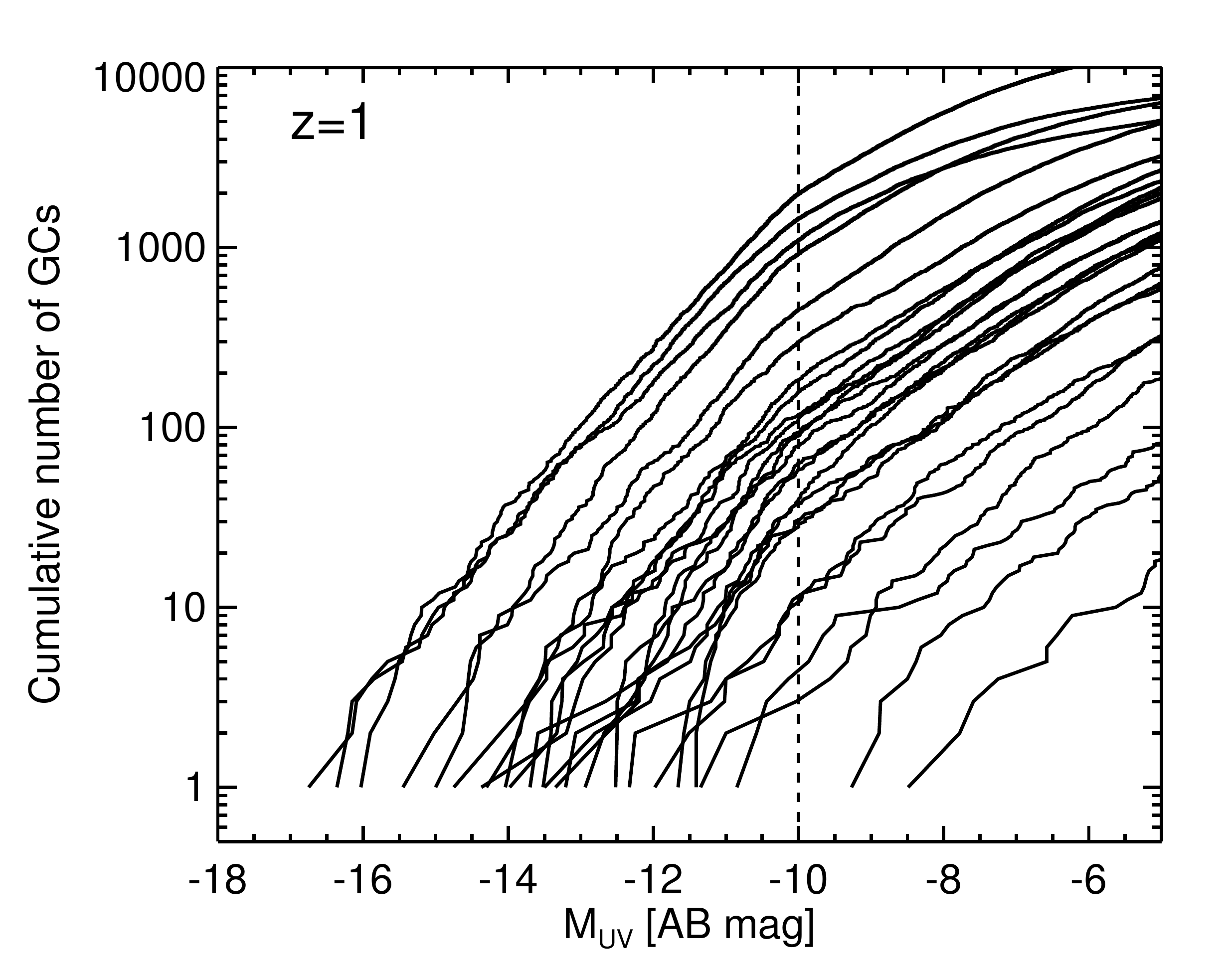}
\includegraphics[width=6cm]{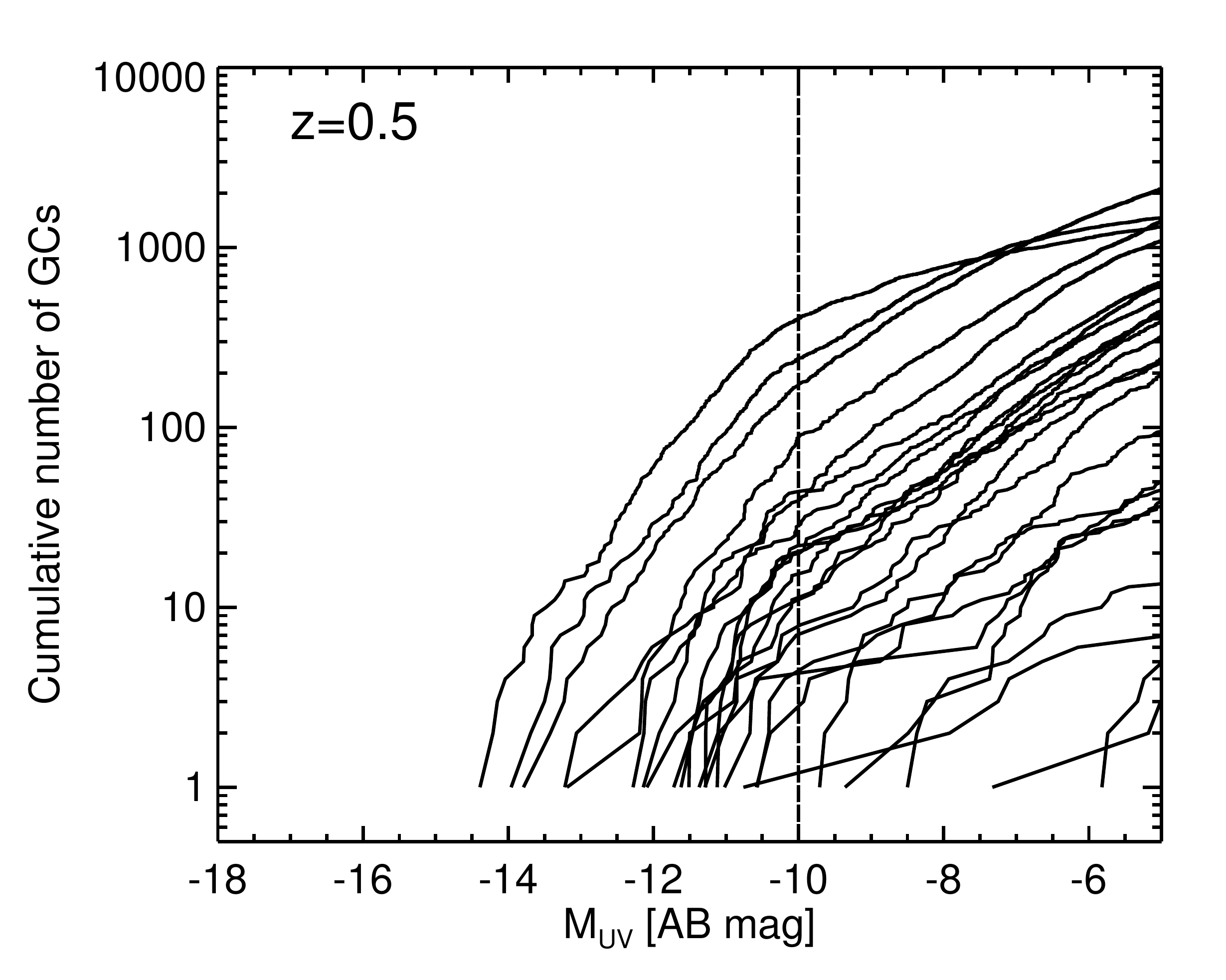}
\includegraphics[width=6cm]{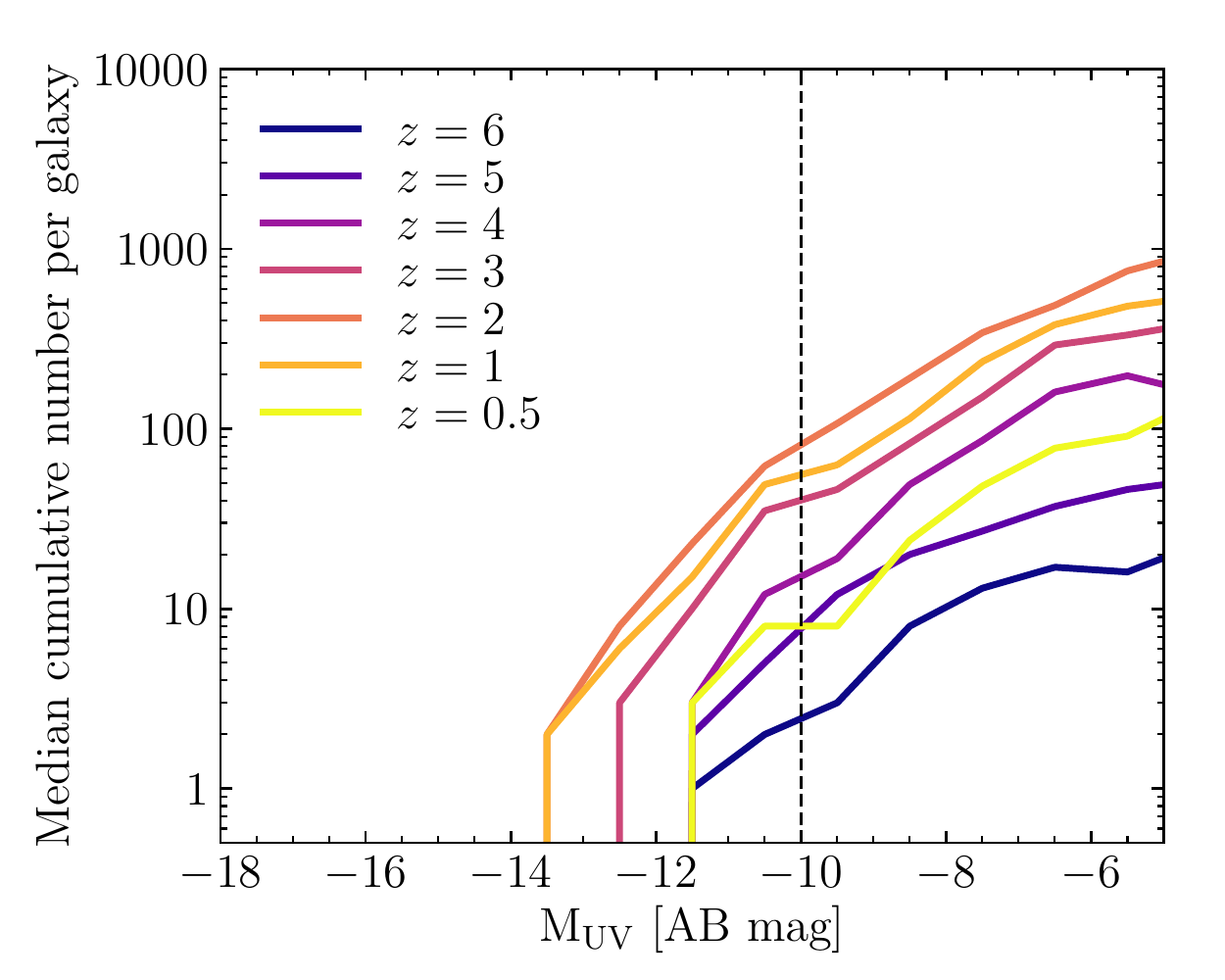}
\caption{The cumulative UV LFs of all 25 haloes at seven redshifts, in the absence of extinction.  The vertical dashed line shows the completeness limit for the adopted lower mass threshold of $5\times10^3 \Msun$.  The scatter between the galaxies is significantly reduced at redshifts 2 and 3 owing to the peak of cluster formation typically occurring around this epoch in the galaxies.  While relatively bright ($\muv<-15$) clusters exist in some galaxies at (nearly) all redshifts, they are most common at redshifts $z=1-3$, reflecting the peak epoch of cluster formation.
The bottom right panel shows the median luminosity function, in bins of $1 \Mag$ width, of all the 25 haloes at each of the seven redshift snapshots. }
\label{fig:lf_cumulative}
\end{figure*}

The cumulative luminosity functions of all 25 E-MOSAICS galaxies are shown in Fig.~\ref{fig:lf_cumulative} for seven epochs. The completeness limit (brighter than which the LF is not affected by the adopted lower-mass limit) is shown as a dashed line at $\muv=-10$.  
In the bottom right panel of Fig.~\ref{fig:lf_cumulative} we show the median GC UVLF for our sample of 25 galaxies for the seven redshift snapshots.
There are various things of interest to note.  
At a fixed detection limit, in the progenitors of Milky Way-mass galaxies we expect to find more young GCs at redshifts $z=1$-$3$ than at other epochs (bottom right panel in Fig.~\ref{fig:lf_cumulative}).
The driver of this evolution is the fact that the cluster formation rate (CFR) peaks at a redshift of $z \sim 2$ in this sample of simulated galaxies \citep{Reina-Campos_et_al_19}.
At these redshifts ($1 \lesssim z \lesssim 3$), about 10--20 per cent of the galaxies host a cluster with $M_{\rm UV}<-15$.
Since the star-formation histories of more massive galaxies are shifted to earlier epochs \citep[e.g.][]{Qu_et_al_17}, we expect the typical cluster formation histories of massive galaxies to be shifted in the same direction \citep[and conversely for less massive galaxies; i.e. galaxy downsizing,][]{Bower_Lucey_and_Ellis_92, Cowie_et_al_96, Heavens_et_al_04, Gallazzi_et_al_05, Nelan_et_al_05}.

\begin{figure}
\centering
\includegraphics[width=84mm]{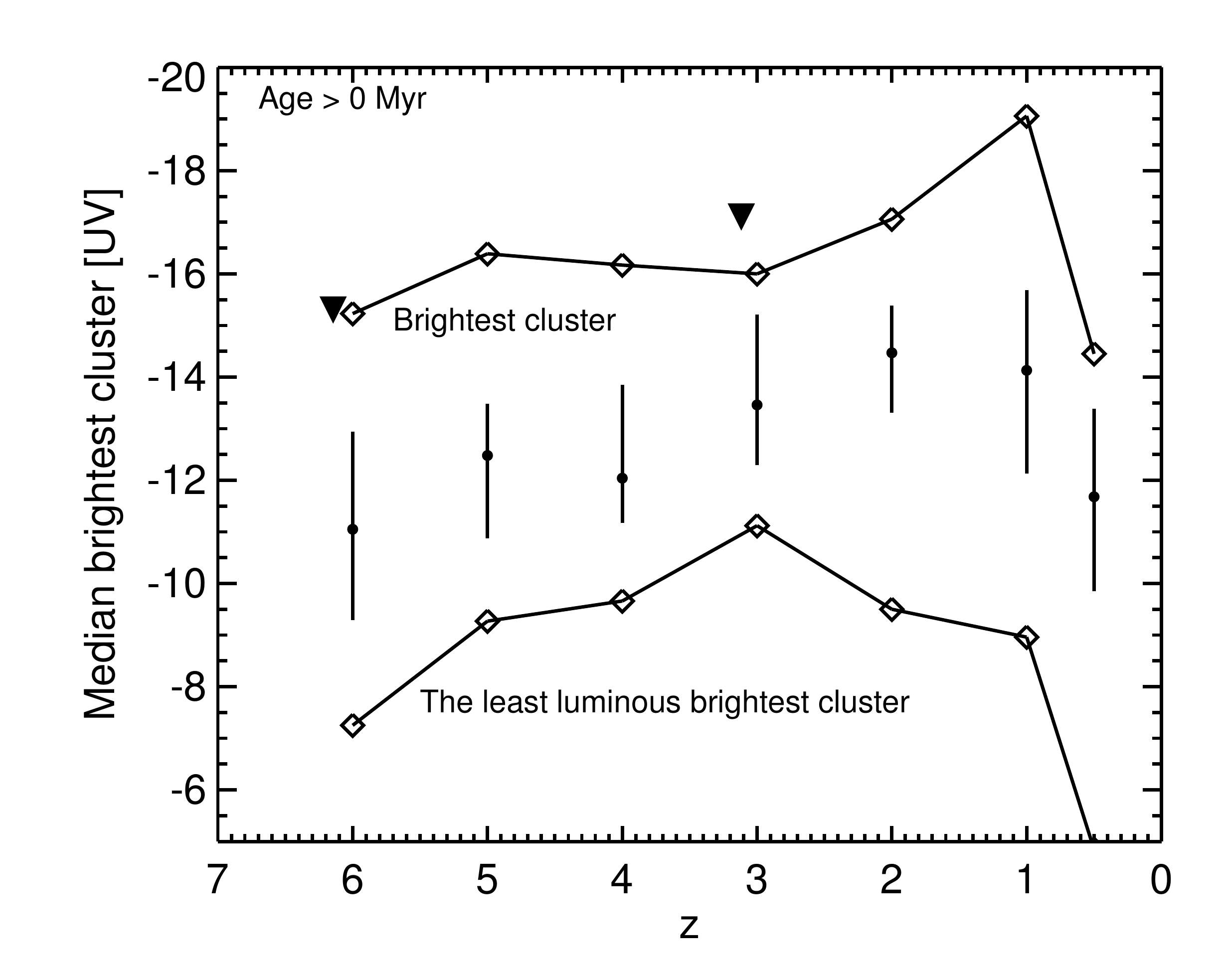}
\includegraphics[width=84mm]{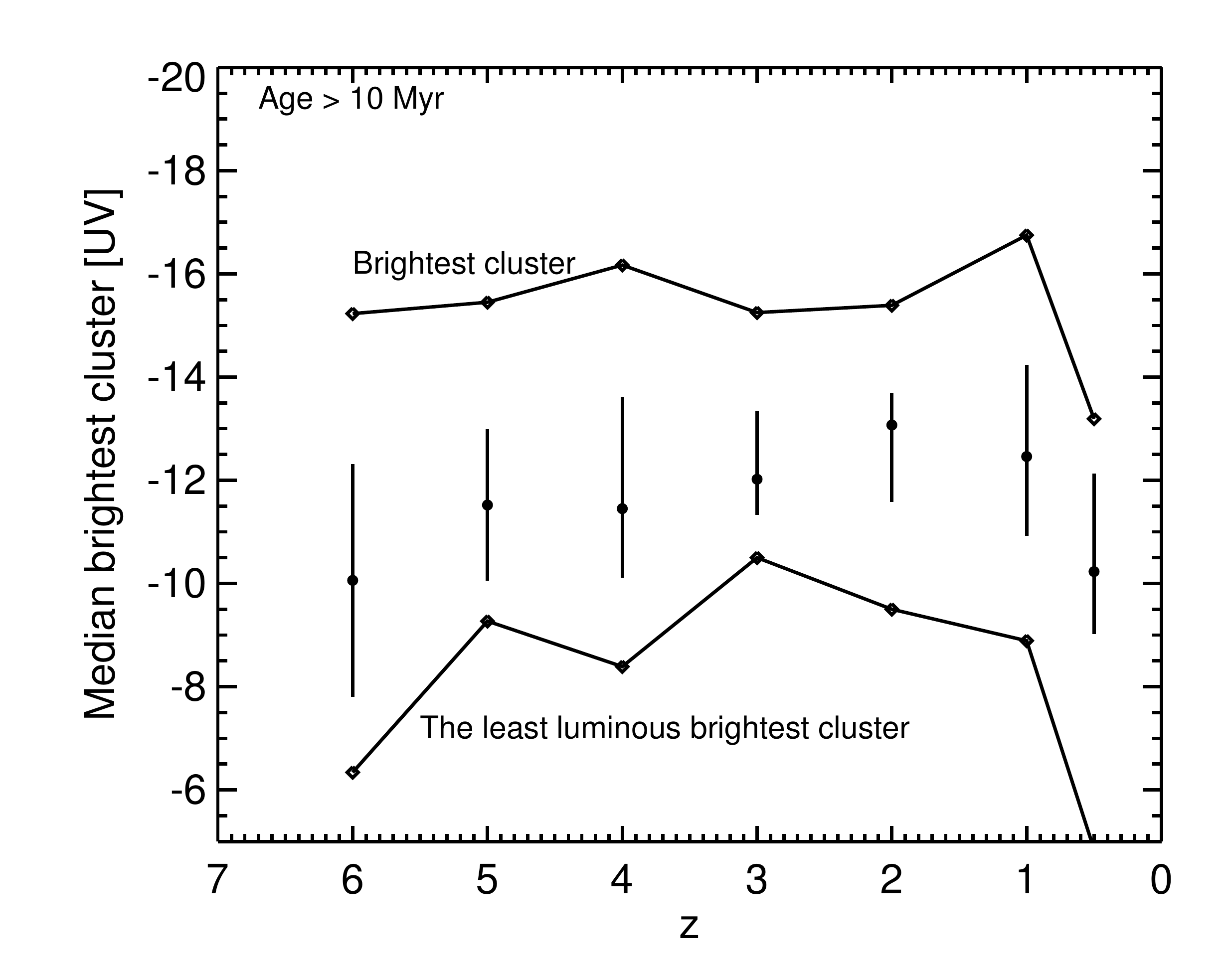}
\caption{{\bf Top panel:} The median brightest cluster (in $\muv$) of all 25 galaxies in our sample (filled dots, no age restriction applied).  The error bars show the $16^{\rm th}$-$84^{\rm th}$ percentiles of the sample at each redshift.  Additionally, we show the brightest cluster in our sample (upper solid line) as well as the least luminous `brightest cluster' in the sample (lower solid line).  For typical current sensitivities of $\muv \lesssim -15$ it is clear that at all redshifts the majority of Milky Way-mass progenitors are not expected to have a young GC detected within them. At redshifts $z=1{-}3$, some 10--20 per cent of the galaxies host a cluster with $M_{\rm UV}<-15$. The upside down triangles denote objects from \citet{Vanzella_et_al_17a} that have measured effective radii $<50$~pc (potentially young GCs). {\bf Bottom panel:} The same as the upper panel but now only considering clusters with ages $>10$~Myr, simulating the effect of all young clusters being heavily extincted.}
\label{fig:uv_scatter}
\end{figure}

For the sample under study in Fig.~\ref{fig:lf_cumulative}, the scatter in the bright end of the LF is very large for redshifts $z>3$ and $z<2$.  However, for a relatively narrow window ($z=2$-$3$) the LFs converge with few outliers.
We quantify this scatter in the top panel of Fig.~\ref{fig:uv_scatter}, where we show the median magnitude and $16^{\rm th}$-$84^{\rm th}$ percentiles (points with errorbars), as well as the maximum and minimum luminosities of the brightest cluster within each galaxy.
The median magnitude of the brightest cluster increases from $\muv=-11$ at $z=6$ to $\muv=-14.5$ at $z=2$, and then decreases towards lower redshifts.  
Hence, given current detection thresholds with HST imaging \citep[$\muv \sim -14$][]{Bouwens_et_al_17c}, the majority of high-redshift progenitors of Milky Way-mass galaxies ($M_{vir} \approx 10^{12} \Msun$ at $z=0$) would not be expected to have individual young GCs detected within them until a redshift of $z \sim 2$, although we predict some detections at any $z\geq1$.
The brightest and faintest clusters (of the brightest cluster sample) follow the same trend as the median, shifted to brighter or fainter magnitudes, respectively.
The $16^{\rm th}$-$84^{\rm th}$ percentiles show the narrowest distribution at $z=2$, where the cluster UV LFs of the galaxies converge. 
This can be attributed to the high CFRs in the galaxies at this epoch. Though the epoch of the peak CFR differs between individual galaxies, at $z=2$ the majority of the Milky Way-mass progenitor galaxies have high ($>0.1 \Msun \yr^{-1}$) and sustained CFRs, such that cluster mass function is continually being well sampled.
At earlier times, the cluster UV LFs are determined by the chance of observing a galaxy when massive clusters happen to be forming, while at later times the galaxies evolve differently in terms of their star and cluster formation rates (i.e. some galaxies become quenched in star formation).

As is also shown in Fig.~\ref{fig:uv_scatter}, the brightest clusters in our galaxy sample are consistent with the luminosities of compact (effective radii $<50\pc$), high-redshift objects detected by \citet{Vanzella_et_al_17a}.
However, it is important to note that the progenitors of our 25 Milky Way-mass galaxies may not be representative of the host galaxies observed at high redshift. 
Direct comparisons between the simulations and high-redshift observations therefore requires comparable galaxy selection criteria, which is beyond the scope of this work.

\subsubsection{With extinction}

We now estimate the effect of extinction on the observed cluster luminosities by exploring an extreme limiting case.
For this we assume a step-function for the extinction, i.e., the cluster is in either an optically thin or optically thick environment.  As clusters are born within larger GMCs and GMC complexes, and emerge once the young cluster destroys its progenitor cloud \citep[e.g.][]{Oort_and_Spitzer_55, Whitworth_79, Larson_81, Murray_11} or migrates away from it \citep[e.g.][]{Kruijssen_et_al_11}, we expect the extinction to be a strong function of its age \citep[e.g.][]{Charlot_and_Fall_00}.  This is observed in young clusters in the local Universe \citep[e.g.][]{Whitmore_et_al_11, Hollyhead_et_al_15, Grasha_et_al_18, Kruijssen_et_al_19c}.  We assume that the clusters are born in an optically thin cloud (essentially all clusters are visible from $t>0$) or become so after some time $\Delta(t$), for which we adopt the (largely) limiting case of $\Delta(t)=10$~Myr.  Due to the strong evolution of the rest-frame UV luminosity of clusters as a function of age, restricting the sample to only clusters older than $10$~Myr drastically affects the bright end of the luminosity function.

This can be seen in the bottom panel of Fig.~\ref{fig:uv_scatter}, which is the same as the top panel, but shows only clusters older than $10$~Myr.  Overall, the population shows the same distribution, however, the brightest clusters are $0.75-1.5$~mags fainter.  Age cuts between $0$ and $10$~Myr give results intermediate between these extremes.

Observations of high-redshift clumps have generally found low extinction values \citep[e.g.][]{Vanzella_et_al_17a}.  However, this is possibly a selection effect as only young, low-extinction, sources are likely to be bright enough to be detectable.

\subsection{Cluster complexes and clumps}
\label{sec:complexes}

In the local Universe, from starburst galaxies to quiescent spirals, stellar clusters rarely form in isolation, but rather do so as part of a larger hierarchy of star-formation within a galaxy \citep[e.g.][]{Zhang_Fall_and_Whitmore_01}.  Young clusters are thus often found as part of larger unbound `cluster complexes', that dissolve within $10-20$ Myr into the surrounding field or halo of the most massive cluster \citep[e.g.][]{Larsen_et_al_02, Bastian_et_al_13, Grasha_et_al_17}. 
In the local Universe, these complexes have effective radii of tens of parsecs, meaning that at high redshift they would often be unresolved, even in highly-magnified HST imaging of gravitationally lensed sources with a resolution of $\sim 50 \pc$ \citep[though some objects have measured sizes of $\sim10 \pc$, approaching that of individual clusters, see][]{Bouwens_et_al_17c, Vanzella_et_al_17a, Vanzella_et_al_17b, Vanzella_et_al_19}. 

Observations of high-redshift galaxies appear to show similar behaviour, with young stars (and ionized gas) being preferentially found in large clumps with the inferred sizes and masses dependent on the resolution of the observations \citep[e.g.][]{Dessauges-Zavadsky_et_al_17, Rigby_et_al_17, Cava_et_al_18}.  Hence, in many cases, observed sources in high-redshift galaxies might not be individual young GCs, but rather part of large clumps or complexes which may significantly increase the inferred brightness of the young GC.

We can use local cluster complexes to estimate the possible scale of such an effect.  \citet{Bastian_et_al_05, Bastian_et_al_06} studied cluster complexes in the nearby spiral galaxy, M51, and the Antennae galaxy merger, respectively.  Since the complexes were (nearly) fully resolved, the authors were able to estimate the fraction of light contributed by the brightest (most massive) cluster within the complex.  They found that each complex was on average $1.5$ to $2.5$ magnitudes brighter than the brightest cluster in the optical.  
However, due to the rapid stellar evolutionary fading in the UV and the increased contribution of unrelated field stars in redder filters, we may expect this effect to be somewhat reduced at bluer wavelengths.

Using UV (F275W) observations of the Antennae galaxies (HST-GO:14593, PI Bastian) we tested the effect of resolution (aperture size) directly.  As input we used the 10 brightest clusters (as estimated in the optical) from \citet{Whitmore_et_al_10}.  We conducted aperture photometry of these clusters, varying the aperture radius from 3 pixels ($\sim12$~pc) to 25 pixels ($\sim100$~pc).  We also measured the corresponding change for unresolved stars in the outskirts of the images, although these tended to be much fainter than the clusters, as well as isolated brighter clusters.  Comparing the magnitude recovered using the $\sim12$~pc aperture and the $\sim50$~pc aperture, we find that the latter was brighter by $0.8\pm0.3$ magnitudes.  The unresolved stars and individual clusters have a difference of $0.3$ mags, showing that the majority of the difference in the cluster sample was due to the inclusion of the surrounding stellar populations.  Comparing the $12$ and $100$~pc apertures yields a difference of $1.1\pm0.5$ mag.

This suggests that the observed luminosities (and derived masses) of the proto-GCs in high-redshift observations might not correspond to a single cluster, but rather to a complex.  If so, the true luminosities of the brightest clusters are likely 0.8-1.1 magnitudes fainter, owing to the additional flux contributed by the surrounding lower-mass clusters and associations.  However, there is significant scatter for individual clusters/complexes.

This notwithstanding, there are clear cases of massive clusters forming (ages $<5$~Myr) essentially on their own (i.e. no other clusters within $\sim50$~pc), even within M51 and the Antennae.  Hence, a universal correction is not advisable, as some observations may identify individual clusters while others may identify complexes, even at the same resolution.
It is likely that the observed clumps in high-redshift galaxies host young GCs, although many clumps are likely to host multiple young GCs rather than only one \citep[e.g.][]{Shapiro_Genzel_and_Forster_Schreiber_10, Kruijssen_15}.  The number density of these clumps may therefore be a good tracer of the cosmic GC formation rate \citep[see also][]{Reina-Campos_et_al_19}. Surveys of such clumps have found that the fraction of `rest-frame UV clumpy galaxies' varies strongly with redshift, with a peak at $z\approx2$ \citep{Shibuya_et_al_16} corroborating the overall GC formation history predicted by E-MOSAICS \citep{Reina-Campos_et_al_19}.

\subsection{The fraction of UV light contributed by GCs}
\label{sec:uv_frac}

\begin{figure}
\includegraphics[width=84mm]{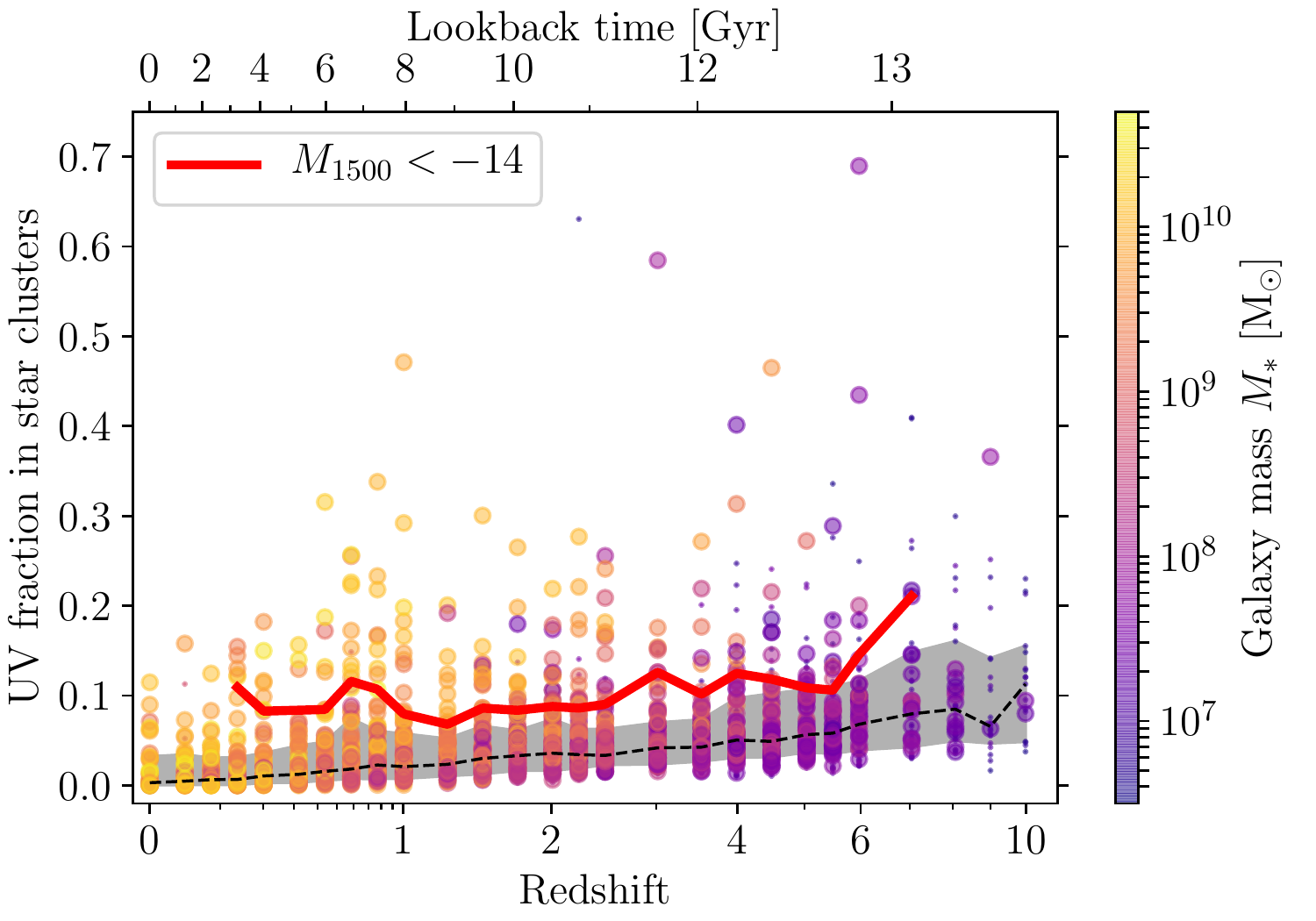}
\includegraphics[width=84mm]{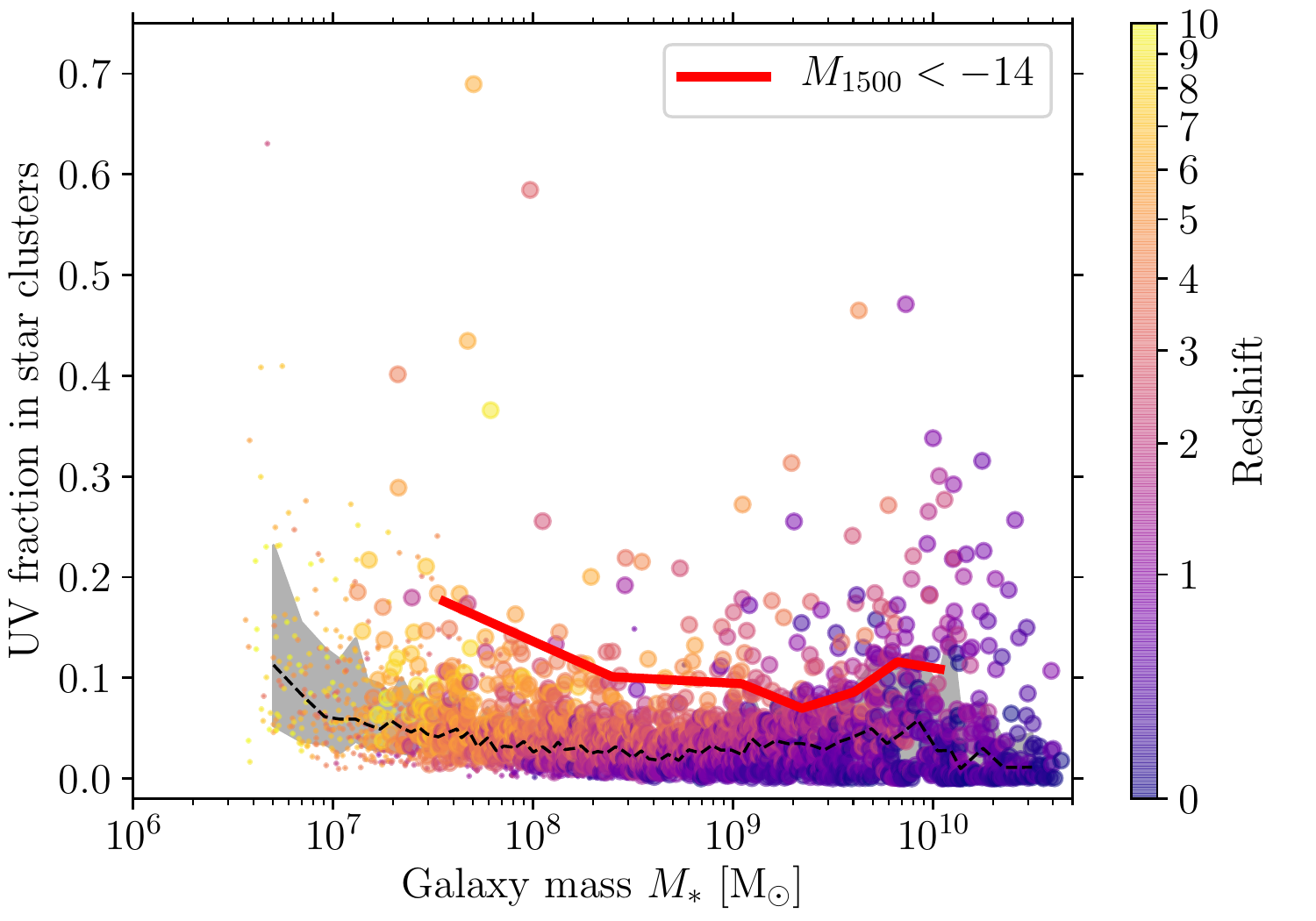}
\caption{ The fraction of UV flux in a galaxy contributed by the full star cluster population for all progenitors of the 25 Milky Way-mass galaxies and their $z=0$ satellites. The top panel shows the UV cluster fraction as a function of redshift (colour-coded by galaxy mass at that epoch), while the bottom panel shows the UV fraction as a function of galaxy mass (colour-coded by redshift). Large points show galaxies with at least 50 star particles younger than $100 \Myr$ (resolved recent star formation histories), small points show galaxies with $20 \leq N < 50$ star particles younger than $100 \Myr$ (partially-resolved recent star formation histories). The dashed black line and the grey shaded region shows the median and $16^{\rm th}$-$84^{\rm th}$ percentiles for all galaxies, respectively (at each redshift in the top panel; in bins of 40 galaxies in the bottom panel). The thick red line shows the median only for galaxies with a cluster brighter than $M_{1500} < -14$. }
\label{fig:uv_frac}
\end{figure}

\begin{figure}
\includegraphics[width=84mm]{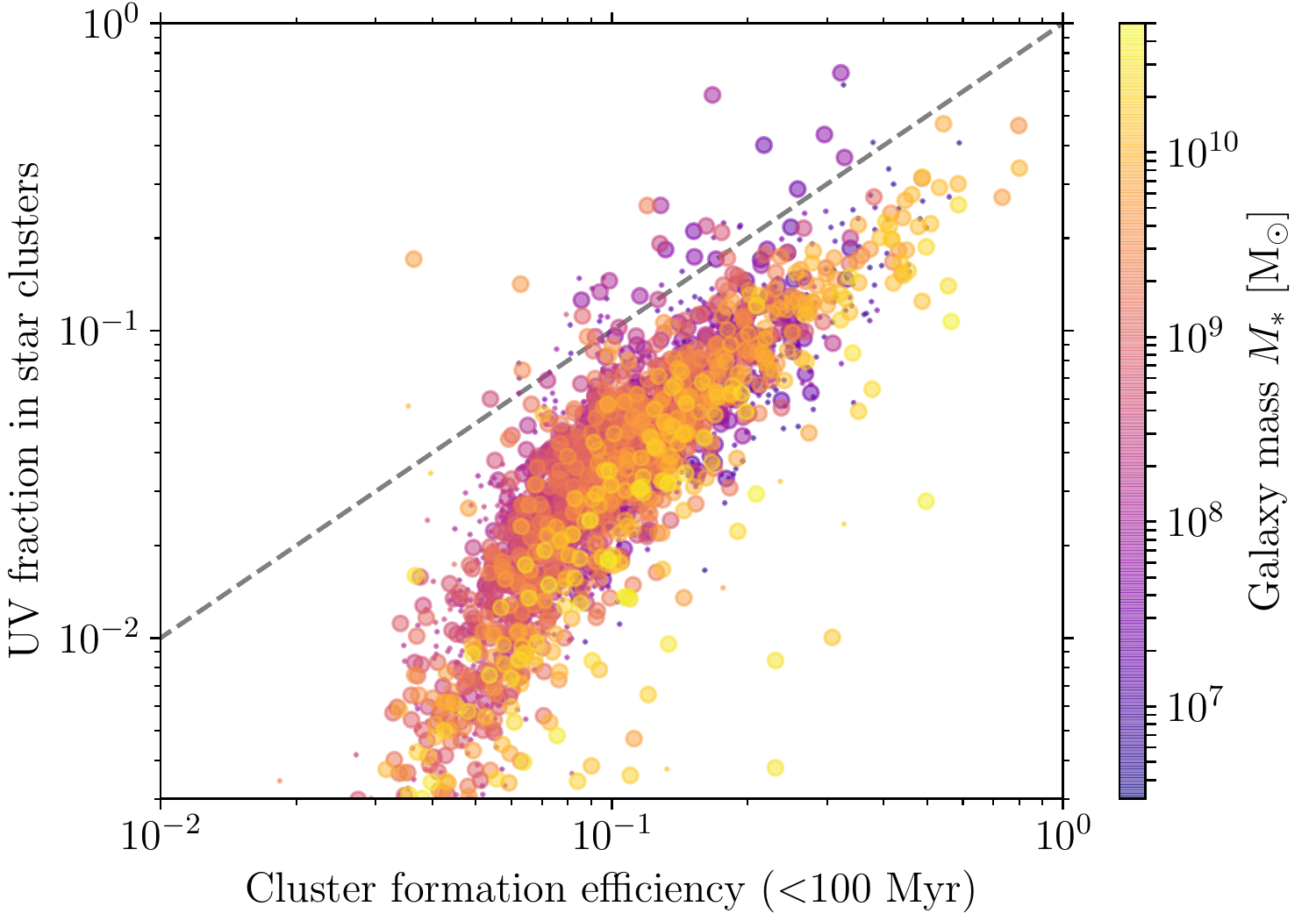}
\caption{ The fraction of UV flux in a galaxy contributed by star clusters compared with the cluster formation efficiency (CFE, calculated for ages $<100 \Myr$) for all progenitors of the 25 Milky Way-mass galaxies and their $z=0$ satellites. Point sizes are as in Fig.~\ref{fig:uv_frac}. The dashed grey line shows the one-to-one relation. In the majority of galaxies, the UV fraction is lower than the CFE (by a typical factor of $0.4$ for galaxies with CFE $>0.1$). The downturn at CFE $<0.1$ is due to the instantaneous disruption of clusters less massive than $5000 \Msun$ in the model. }
\label{fig:uv_frac_CFE}
\end{figure}

\begin{figure}
\includegraphics[width=84mm]{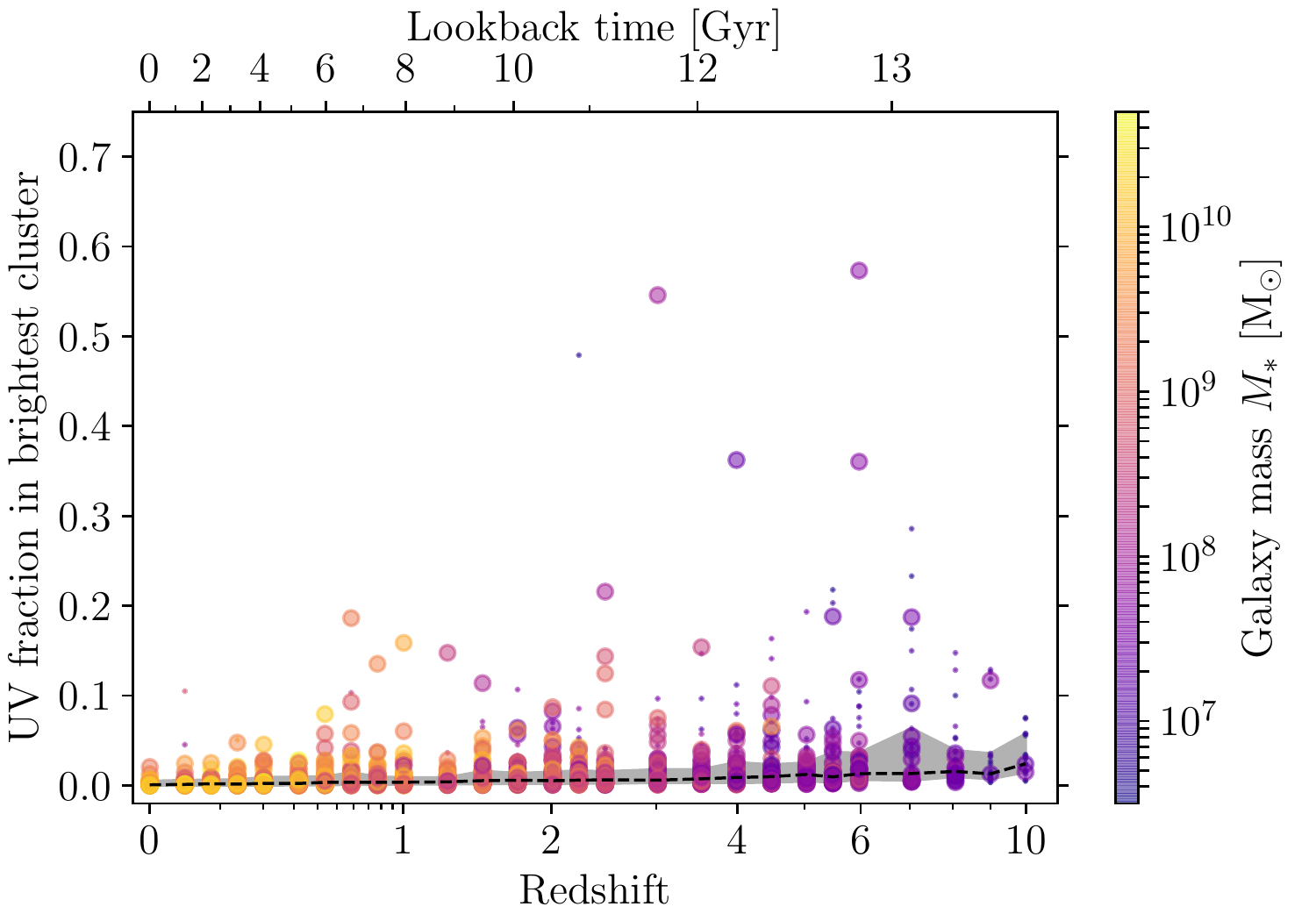}
\includegraphics[width=84mm]{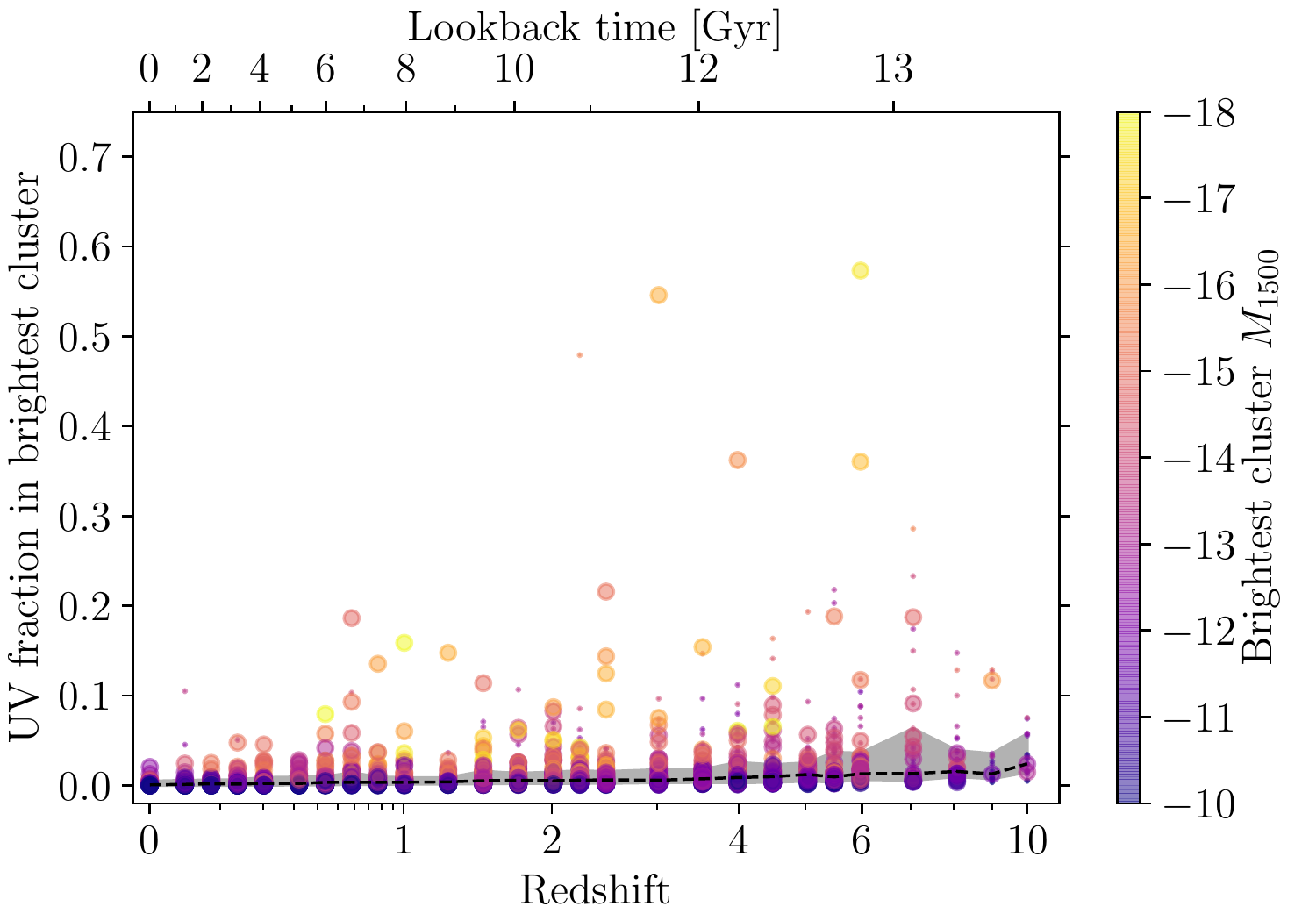}
\includegraphics[width=84mm]{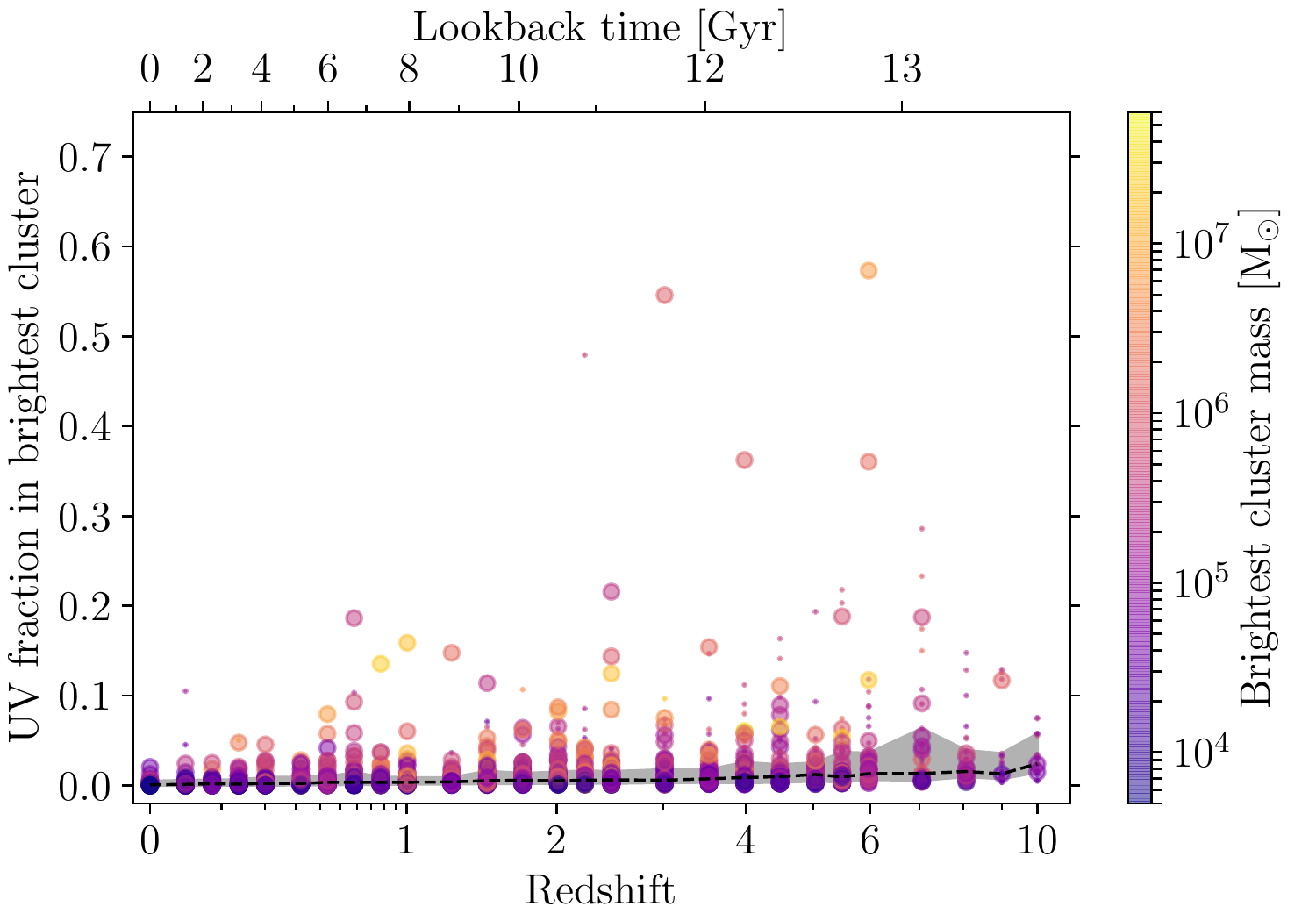}
\caption{ The fraction of UV flux in a galaxy contributed by the brightest cluster in the galaxy for all progenitors of the 25 Milky Way-mass galaxies and their $z=0$ satellites. Symbols are colour-coded by the galaxy stellar mass (top panel), UV luminosity of the brightest cluster (middle panel) and mass of the brightest cluster (bottom panel). The dashed black line and the grey shaded region shows the median and $16^{\rm th}$-$84^{\rm th}$ percentiles for all galaxies, respectively. Point sizes are as in Fig.~\ref{fig:uv_frac}. }
\label{fig:uv_frac_brightest}
\end{figure}

In this section, we investigate the fraction of the UV luminosity of a galaxy contributed by star clusters. 
To do this, we apply the same method of calculating the brightness of the clusters to the field stars of each star particle within the simulations \citep[see also][]{Trayford_et_al_15}. 
Briefly, we assign each star particle a luminosity based on their age, mass and metallicity (assuming a simple stellar population) following the method described in Section \ref{sec:SSPs}.
At each redshift, we sum the total UV flux in stars and either sum the total flux in clusters, or consider the brightest cluster in the UV at that epoch.  
For these tests we will assume the case of no extinction.

We model star formation by converting entire gas particles into stellar particles, therefore the mass resolution of the simulations imposes a sampling limit on star formation histories, i.e. each star formation episode results in the formation of at least $2\times10^5~\Msun$ of stars.
In principle, this could lead us to overestimate the relative UV brightness of stellar populations. For this effect to be important, star-forming regions would need to have age spreads in excess of the duration of the UV-emitting phase, which lasts $15{-}35$~Myr \citep[e.g.][]{Haydon_et_al_19}. 
However, observations of molecular clouds and star-forming regions in the local Universe show that molecular clouds are dispersed within $1{-}6$~Myr after the emergence of the first massive stars \citep{Kruijssen_et_al_19c,Chevance_et_al_19}, much shorter than the time for which young stellar populations are UV-bright. 
This means that the star formation histories of clusters in the local Universe are effectively delta functions. 
Even though these observations only consider nearby galaxies, they span a wide variety of galactic environments, with gas surface densities in the range $\Sigma=1{-}200~\Msun~\pc^{-2}$. 
We therefore expect these conclusions also apply under the conditions of GC formation.

In Fig.~\ref{fig:uv_frac} we show the results for all 25 Milky Way-mass haloes.  
The figure includes all progenitors of the main galaxy, as well as those of the galaxies identified as satellites at $z=0$. 
We only include galaxies with at least 20 stellar particles younger than $100 \Myr$ (implying a minimum SFR of $\approx 0.04 \Msun \yr^{-1}$), i.e. those galaxies with a reasonably well sampled recent star formation history.
This limit therefore implies, at a fixed specific SFR, better sampling of the SFR in higher mass galaxies.
In the top panel of Fig.~\ref{fig:uv_frac}, we show the UV fraction as a function of redshift. Note that where the temporal resolution between snapshots is shorter than a few hundred megayears, the measurements for galaxy descendents/progenitors may not be independent (see also Fig.~\ref{fig:age_uv}).
For all galaxies, the median UV fraction of star clusters in galaxies decreases from 10 per cent at $z=10$, to 0.3 per cent at $z=0$, with a median for all galaxies of $3.5$ per cent. 
This fraction only varies mildly with galaxy stellar mass (bottom panel), with a small upturn to a UV fraction of $\sim10$ per cent at galaxy masses $\lesssim 10^7 \Msun$.
If we consider only galaxies where a star cluster could be detected ($M_{1500} < -14$), the UV fraction in clusters is typically $\sim10$ per cent at all epochs (solid red line in the top panel).
Therefore, for a typical galaxy observed at high redshift, field stars should always dominate the UV light.
A similar result, where clusters contribute $\lesssim 50$ per cent of the UV flux, is found for young clusters in local Universe (e.g. \citealt{Larsen_and_Richtler_00}; see \citealt{Adamo_and_Bastian_18} for a review), which is expected given the cluster formation model in E-MOSAICS is based on young clusters.

The subdominant fraction of UV flux contributed by clusters can be explained in the model by considering the CFE near the time of the snapshot, which we compare against the UV fraction in Fig.~\ref{fig:uv_frac_CFE}. The CFE is calculated for a galaxy by summing the total mass in clusters formed for all star particles with ages $<100 \Myr$.
In general, the UV fraction in a galaxy is always less than the CFE, and typically a factor of 0.4 lower (for galaxies with CFE $>0.1$), due to the fading of clusters older than $\sim 10 \Myr$ (see Fig.~\ref{fig:age_uv}).
A small fraction of galaxies have a UV fraction larger than the CFE. These outliers tend to be low-mass galaxies ($<10^8 \Msun$) which happen to host a single (or a few) young, bright clusters (see Fig.~\ref{fig:uv_frac_brightest} and discussion below).
For more massive galaxies, a high UV fraction implies a high CFE, since such galaxies have well sampled star and cluster formation rates. 

In a few cases, which are typically low-mass galaxies with stellar masses $M_\ast<10^8 \Msun$, clusters can dominate the light of the galaxy, reaching a peak fraction of $\approx0.7$. 
However, such galaxies are rare: just $0.3$ per cent of galaxies with $M_\ast < 10^8 \Msun$ at $z\geq2$ (similarly for $z\geq4$) have a UV fraction $>0.5$ (or $0.2$ per cent of all galaxies at $z\geq2$).
In Fig.~\ref{fig:uv_frac_brightest} we show the UV fraction contributed by the single brightest cluster in the galaxy. 
The behaviour is qualitatively similar to that of the full population. The typical contribution of the brightest cluster to the UV flux in a galaxy is $\approx 0.5$ per cent, which is relatively independent of redshift.
Many of the low-mass galaxies with high total UV fractions (Fig.~\ref{fig:uv_frac}) also retain high UV fractions (up to $\approx 0.6$) when only considering the brightest cluster, since the cluster UV flux is being dominated by a single object.
In the six galaxies where the UV fraction of the brightest cluster is $>0.25$, the brightest clusters have UV luminosities $M_{1500} < -15$ (middle panel of Fig.~\ref{fig:uv_frac_brightest}), and would therefore be readily detectable in current gravitational lensing surveys. 
The clusters of these galaxies all have masses $\gtrsim 5 \times 10^5 \Msun$ (bottom panel of Fig.~\ref{fig:uv_frac_brightest}) and ages $< 10 \Myr$.
In these six galaxies, the brightest cluster contains $\sim 10$ per cent (or less) of the galaxy stellar mass, so although it contributes a large fraction of the UV flux, it never dominates the mass. 
In terms of the mass fraction in clusters, these galaxies may potentially be similar to some local dwarf galaxies (Fornax, WLM, IKN), where GCs account for $\sim$20-30 per cent of the mass fraction of low metallicity stars \citep{Larsen_Strader_and_Brodie_12, Larsen_et_al_14}.

\begin{figure}
\includegraphics[width=84mm]{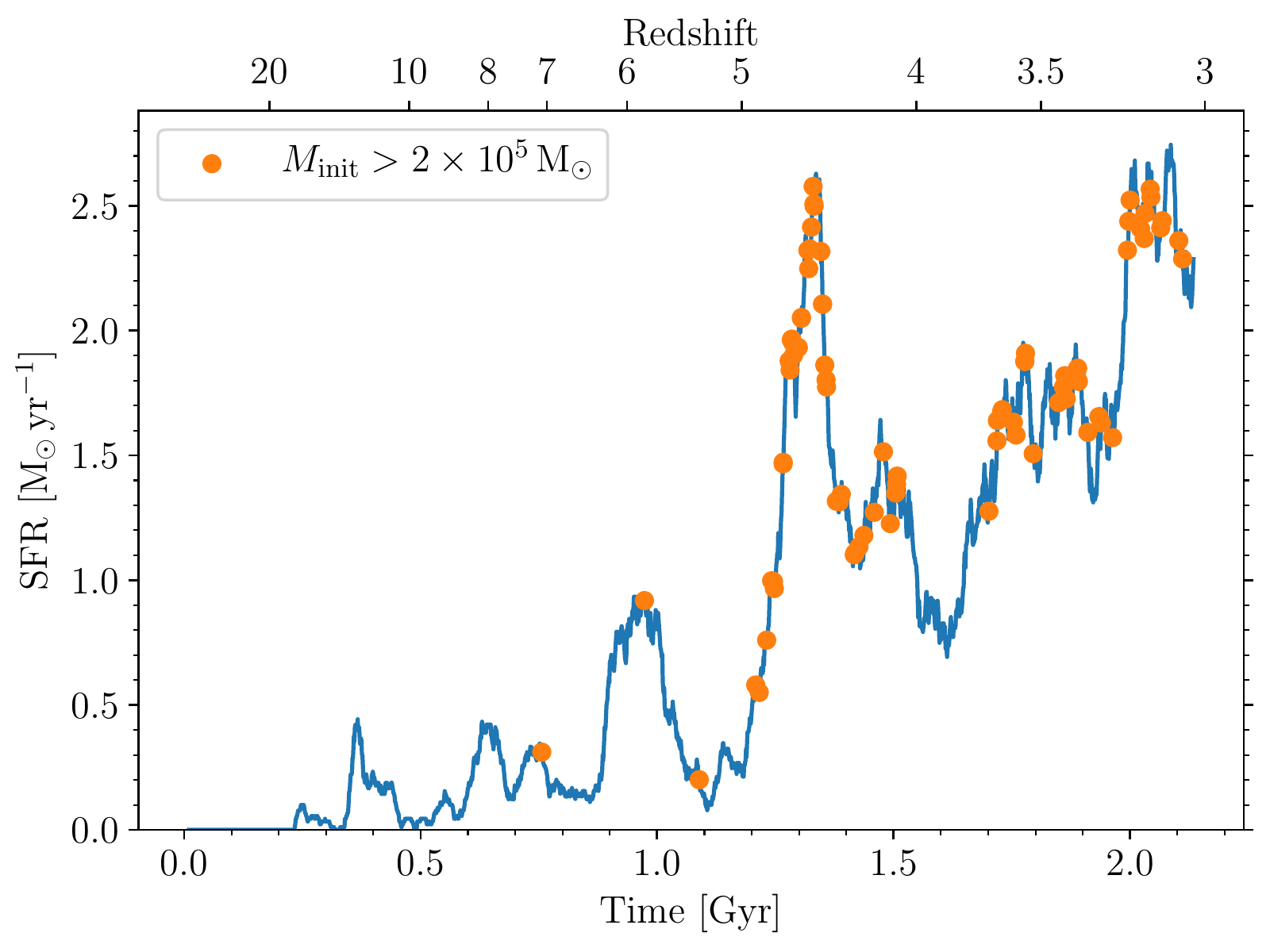}
\caption{Formation times of massive clusters (with initial masses $>2 \times 10^5 \Msun$) relative to the galaxy SFR for MW00 at redshifts $z>3$ (reconstructed from the $z=3$ snapshot). Massive clusters preferentially form at the (relative) peaks of the SFR, explaining why clusters almost never dominate the total UV flux of a galaxy (Fig.~\ref{fig:uv_frac}).}
\label{fig:formtime}
\end{figure}

Due to the rapid fading, the UV fraction is mostly governed by the (instantaneous) value of the CFE, i.e. the fraction of star formation that takes place within bound clusters, at any given epoch.  Even following a period of a starburst (and associated large CFE values) the UV fraction rapidly declines to the new value set by the current value of the CFE.
Additionally, the formation of massive clusters is biased to periods of high SFRs, when the pressures and densities of star-forming gas are high, resulting in more efficient cluster formation \citep[see also the discussion in][]{P18}.
In Fig.~\ref{fig:formtime}, we show the formation times of massive clusters (with initial masses $>2 \times 10^5 \Msun$) relative to the SFR of the galaxy MW00, where the SFR is reconstructed from the $z=3$ snapshot. Therefore the SFR includes any galaxies which have merged prior to $z=3$ (of which eight mergers occurred with galaxies with at least 20 stellar particles, i.e. $M_\ast \gtrsim 10^{6.5} \Msun$). Note that cluster formation appears to occur earlier in Fig.~\ref{fig:age_uv} due to the cluster mass limit imposed in Fig.~\ref{fig:formtime}.
The formation of massive clusters is biased to the (relative) peaks in the SFR, and thus clusters preferentially form when the galaxy is expected to be UV-bright. This explains why clusters typically never dominate the UV flux in a galaxy, even when it hosts bright clusters (Figs.~\ref{fig:uv_frac} and \ref{fig:uv_frac_brightest}).

Similar results have also been investigated in the literature by other authors.
\citet{Zick_Weisz_and_Boylan-Kolchin_18} reconstructed the evolution of the UV luminosity of the Fornax dwarf spheroidal galaxy and its GC population, based on their current properties.  The authors focus on the fraction of UV light emitted by the young GCs relative to the field population of the host galaxy.  By stochastically adding GC formation onto the inferred UV luminosity of the field, based on the best-fitting star formation history and assumed SFR modulation over time, the authors conclude that GCs can contribute $>95$ per cent of the UV light, even though they comprise $<5$ per cent of the stellar mass.  The main underlying assumption of the work is that the formation of field stars and GCs is not correlated.  
Such high UV fractions are not found in our simulations due to the assumed models for the CFE and cluster mass function (based on young cluster populations in the local Universe). In the model, star and star cluster formation proceed in a correlated fashion (Fig.~\ref{fig:formtime}), resulting in a UV fraction generally below 10 per cent (Fig.~\ref{fig:uv_frac}).
Even in low-mass galaxies ($M_\ast <10^8 \Msun$) at $z>2$, galaxies where the UV fraction is higher than 50 per cent represent just $0.3$ per cent of the population of progenitors of Milky Way-mass galaxies and their satellites.

\subsection{Detecting GCs and their host galaxies at high redshift}
\label{sec:sb}

\begin{figure*}
\includegraphics[width=\textwidth]{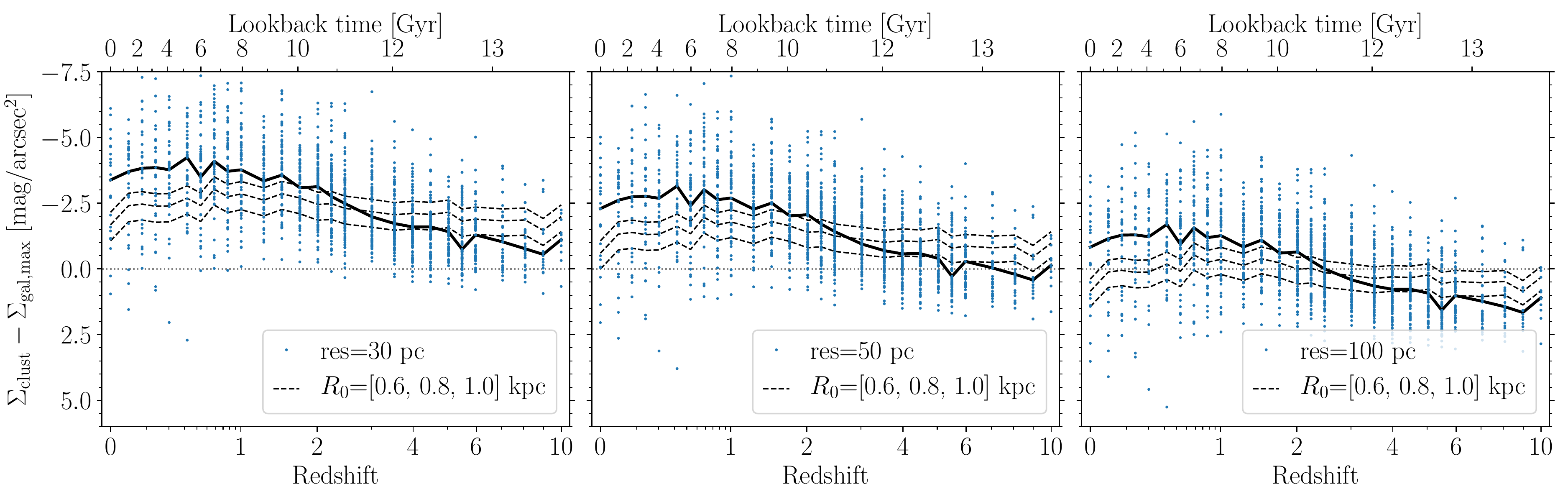}
\caption{The difference in surface brightness in the rest-frame UV of the brightest cluster and the brightest pixel in the galaxy at resolutions of $30\pc$ (left), $50\pc$ (middle) and $100\pc$ (right). The solid line shows the median value at each snapshot.  The sizes of the galaxies are scaled to evolve with redshift and galaxy mass (see text) and the three panels show the effect of resolution. The dashed lines show the median values for sizes that evolve with redshift but are independent of mass (see text), with larger sizes leading to brighter clusters relative to the galaxy (see the legend).  
At higher resolution, the brightest clusters show more contrast with the host galaxy, though clusters are typically not significantly brighter than the peak surface brightness of the galaxies. 
  }
\label{fig:sigma}
\end{figure*}

Despite a considerable difference in luminosity, young GCs are however expected to exhibit surface brightnesses much higher than their much more extended host galaxies due to their compactness.  Based on this, in combination with their rapid fading in the rest-frame UV, it is conceivable that individual GCs are observed (with the host galaxy undetected in the observations as well as other young GCs) and mistaken for compact galaxies \citep[e.g.][]{Bouwens_et_al_17c}.

We investigate this possibility in our simulations by calculating, for a given resolution, the rest-frame UV surface brightness of the brightest GC at a given snapshot as well as the surface brightness profile of the host galaxy.  We do this for a variety of spatial resolutions. The UV fluxes are estimated in the same way as above.  To compute the galaxy surface brightness profiles, we do not use the spatial distribution of the field stars from the E-MOSAICS simulations.  
The EAGLE simulations are able to reproduce the mass and star-formation rates of high-redshift galaxies \citep{Furlong_et_al_15} and the size evolution of massive galaxies \citep{Furlong_et_al_17}, but over-predict the spatial extent of the smallest ($<1 \kpc$) galaxies\footnote{Note that since the star-forming discs of the lowest mass galaxies are non-self-gravitating, the gravitational softening length plays no role in setting galaxy sizes \citep{Benitez-Llambay_et_al_18}.} due to the temperature floor of the polytropic equation of state (8000 K at a density $n_\rmn{H} = 0.1 \cmcubed$) imposed on the unresolved star-forming interstellar medium \citep{Schaye_and_Dalla_Vecchia_08}. 
We therefore set the spatial profiles of the galaxies according to observations of high-redshift galaxies.
We assign the integrated UV flux of the galaxy to be distributed according to a \citet{Sersic_63} profile with an index $n = 1.5$ \citep[as in][]{Holwerda_et_al_15}. We also investigated indices of 1 and 2.5, finding qualitatively similar results. 
The sizes of the galaxies then scale with redshift and galaxy mass according to $R_\rmn{e} = R_0(z) (M_\ast/10^9 \Msun)^{\beta(z)}$, where the intercept $R_0$ and slope $\beta$ vary with redshift \citep{Holwerda_et_al_15}. Fitting functions to the data in table 4 from \citet{Holwerda_et_al_15}, we find $R_0(z)/\rmn{kpc} = 2.57 \exp(-0.33 z) + 0.42$ and $\beta(z) = -0.01 z + 0.24$.
As the size measurements of galaxies at high redshift may be somewhat uncertain, we also investigate a size relation that varies only with redshift according to $R_\rmn{e} \propto R_0 (1+z)^{-0.8}$, where the size intercept $R_0 = [0.6,0.8,1.0] \kpc$ at $z=4$ \citep[i.e. the lower-luminosity galaxy relation from][]{Holwerda_et_al_15}.
For each of the 25 simulated galaxies and their progenitors (i.e. all galaxies in the merger tree, and only those with at least 20 stellar particles) we calculate at each snapshot the peak UV surface brightness of the galaxy and the brightest cluster (assuming clusters have sizes much smaller than the resolution) at resolutions of $30$, $50$ and $100 \pc$. 

In the context of highly-magnified gravitationally-lensed sources at high redshifts ($z\gtrsim3$), this analysis assumes perfect reconstruction of the images. Therefore, our analysis does not capture the biases or limitations of lensing observations, such as uncertainty in the lens model \citep[e.g.][]{Meneghetti_et_al_17}, blending with foreground galaxies and intra-cluster light or the effect of shear in extended sources \citep[see][]{Oesch_et_al_15, Bouwens_et_al_17a}. Blending and regions of high shear are both expected to reduce the completeness of observed sources. Including such effects requires `observing' the simulated galaxies in a lensed framework, which is beyond the scope of this work.

Fig.~\ref{fig:sigma} shows the difference in UV surface brightness between the brightest cluster and the brightest pixel in the galaxy for the three resolutions. 
Negative values correspond to the clusters being more readily detectable than their host galaxy.
The difference between the maximum cluster and peak galaxy surface brightness shows a strong relation with redshift, with clusters being most detectable at low redshifts. 
As the typical maximum cluster surface brightness in star-forming galaxies remains approximately constant with redshift (due to the rapid fading of clusters in the UV and the low chance of observing very massive young clusters, such that young low mass and older high mass clusters have similar luminosities; see Fig.~\ref{fig:age_uv}), this trend is driven by the evolution of the peak UV surface brightness of the galaxies as their masses and sizes increase with time. 
The detectability of clusters also depends strongly on the resolution of the observation, meaning that at high resolution ($30 \pc$, left panel) clusters are generally more readily detectable than the galaxy, while at low resolution ($100 \pc$, right panel) the converse is true.
This is caused by the decreasing surface brightness of clusters at lower `observation' resolution since, for the resolutions investigated here, the peak surface brightness of the galaxy is largely insensitive to resolution.
The surface brightness difference shows a much tighter relation with redshift than the galaxy or maximum star cluster surface brightnesses individually. 
This is due to the correlation in star and star cluster formation rates at high redshift (Fig.~\ref{fig:formtime} and \citealt{Reina-Campos_et_al_19}) as clusters preferentially form when the galaxy is UV-bright.

The dashed lines in Fig.~\ref{fig:sigma} show the result when assuming galaxy sizes that only scale with redshift and excluding the scaling with galaxy mass. Between size intercepts ($R_0$) of $0.6$ and $1 \kpc$, the relative brightness of clusters increases by $-1 \Mag \Arcsec^{-2}$. Thus, the surface brightness difference between clusters and the galaxies is somewhat sensitive to the actual sizes of the galaxies.
However, for reasonable galaxy sizes the general predictions remain the same, with clusters becoming progressively more challenging to detect at higher redshifts (for $z \gtrsim 1$).

The declining sizes of galaxies at higher redshifts leads to the result that, even at resolutions of a few tens of parsecs, on average clusters will not be significantly more detectable than the galaxies themselves as observations push to the more-distant Universe.
At $z>4$, clusters are typically only $\approx -1 \Mag \Arcsec^{-2}$ brighter than the galaxy. 
However, there is significant scatter between different galaxies, such that some clusters are up to $-5 \Mag \Arcsec^{-2}$ brighter than their host galaxies, with even larger differences at lower redshifts.
Therefore, in some cases only the cluster may be observable, depending on the surface brightness limits of the observations.

A potential caveat and source of uncertainty in this analysis is the assumption that the UV flux of the galaxy can be approximated by a smooth distribution. 
The surface brightness of the galaxy may decrease further if much of the UV flux of the galaxy is located in clumps/cluster complexes \citep[e.g.][]{Shibuya_et_al_16}.
Additionally, as discussed in Section \ref{sec:complexes} \citep[see also][]{Bouwens_et_al_17c, Vanzella_et_al_17b}, observations at high redshift may be detecting cluster complexes rather than individual clusters. 
If the resolution of observation is much larger than the cluster size, such that both the cluster(s) and complex are unresolved, then the cluster complex will also contribute to the observed surface brightness (see the analysis on the effect complexes on total luminosity in Section \ref{sec:complexes}).
Therefore, the results presented in Fig.~\ref{fig:sigma} give a lower limit to the surface brightness difference between the main galaxy and young star-forming regions within them, since cluster complexes may be significantly brighter than individual clusters.

\section{Discussion}
\label{sec:discussion}

\subsection{Interpreting high-redshift observations}
\label{sec:interpreting}

A number of recent works have attempted to use observations of high-redshift sources to place constraints on GC formation.  In particular, many have attempted to test scenarios for the formation of multiple populations within GCs that assume large mass loss rates (i.e. that GCs were $10-100$ more massive at birth than they are at present).

\citet{Renzini_17} suggested that observations of young GCs in lensed galaxies may provide a strong constraint on how much mass loss a cluster has experienced.  If, as assumed by some models for the origin of multiple populations within GCs, GCs were $>10-100$~times more massive at birth than at present (specifically when the clusters were younger than a few hundred Myr), then they should be brighter by 2.5 to 5 magnitudes than in the absence of such extreme mass loss.  Mass loss at the lower end of this spectrum (being a factor $10$ more massive at birth than at $z=0$) is likely to be difficult to distinguish from uncertainties in the contribution of the surrounding complex as well as the GC age.  However, the other extreme (being a factor $100$ more massive) should be more readily falsifiable.  
From Fig.~\ref{fig:lf_cumulative} (bottom right panel), adding $-5$ magnitudes would move the bright end of the median LF to $<-16$ for all redshifts and $<-18$ for $z=1$-$2$. 
These luminosities are accessible to current facilities, and their detection would imply that the progenitors of most Milky Way-mass galaxies should host at least one observable cluster.  This scenario thus appears at odds with extant constraints \citep[]{Vanzella_et_al_17a, Bouwens_et_al_17c, Boylan-Kolchin_18}.

As discussed in Section \ref{sec:complexes}, in the local Universe, massive clusters do not generally form in isolation, but rather as part of a larger cluster complex. 
If star formation processes are similar at all redshifts, then the clumps seen in high-redshift observations can be considered analogues of local cluster complexes.
The surrounding stellar population can artificially increase the inferred luminosity of the bright/most massive young GC in the clump by $0.8\pm0.3$~mag in the UV (and significantly more in the optical, based on nearby cluster complexes) at an aperture radius of $50 \pc$, similar to the resolutions achievable in highly-magnified HST imaging \citep[e.g.][]{Bouwens_et_al_17a, Vanzella_et_al_17a}.  Additionally, the large age uncertainties in most observations correspond to factors of 10 or more in estimates of the mass from the UV \citep{Pforr_Maraston_and_Tonini_12}.  The result of this is that, other than for the most extreme mass loss models, observations of high-redshift galaxies in the UV are not likely to be able to place a strong constraint on the mass lost by young GCs \citep[however, this is not the case when masses can be derived from spectral energy distribution fitting, e.g.][]{Vanzella_et_al_17b}.

Moving to redder filters (i.e. with JWST) can help, although the effects of the surrounding complexes can become much worse.  The main benefit of moving towards rest-frame optical colours is that it lessens the strong bias towards finding only the youngest GCs.  In many cases, the surrounding complex is expected to dissolve in the field on $10-30$~Myr timescales, meaning that individual clusters may be measured.  The cluster luminosity function of each galaxy, measured in the $g$-band, is very similar to that observed in the UV, just shifted by $\sim1$~magnitude to fainter luminosities.

In either the UV or the optical rest frame, resolutions of $<10-20$~pc are required to mitigate the effect of the surrounding complex.  If these resolutions are achievable for statistically significant samples of Milky Way-mass progenitor galaxies (as anticipated for the upcoming generation of 30m-class telescopes), it should be possible to directly test the model presented in this paper.  We expect to see the most young GCs at redshifts between 1 and 3, although with a significant number of bright young clusters at higher redshift and significant galaxy-to-galaxy scatter.

As discussed above and by \citet{Shapiro_Genzel_and_Forster_Schreiber_10}, GCs are expected to be forming within the large UV-bright clumps observed in high-redshift galaxies.  \citet{Guo_et_al_15} have estimated the fraction of ``clumpy galaxies'' in the HST CANDELS fields as a function of redshift (between 0 and 3) and found that for present day Milky Way mass galaxies there is a peak in the distribution between $z=2$-$3$.  A similar analysis was done by \citet{Shibuya_et_al_16} who found a stronger peak in the UV clumpy fraction at $z=1.5-2$.  
Our models are compatible with these observational constraints, in that we find that progenitors of Milky Way-mass galaxies at these redshifts host the largest number of young clusters \citep[see][]{Reina-Campos_et_al_19}, which we expect will correlate with the number (and fraction) of UV clumps in galaxies.

Other recent works have also investigated the detectability of clusters relative to their host galaxy at high redshifts. 
\citet{Zick_Weisz_and_Boylan-Kolchin_18} concluded that young GCs may be $\sim -14 \Mag \Arcsec^{-2}$ brighter than the galaxy itself (for a Fornax dwarf spheroidal-like galaxy at $z=3$). 
However, in their analysis cluster formation times are not correlated with the galaxy SFR, such that GC formation might occur during the minima of the SFR, when the galaxy is faint in the UV.
In this work we find that, at resolutions typically achieved by high-redshift lensing studies, young GCs in the progenitors of Milky Way-mass galaxies and their satellites typically do not have significantly higher surface densities than their host galaxies (Fig.~\ref{fig:sigma}).
This difference is caused by the correlation between the formation times of massive clusters and the SFRs of the galaxies in the E-MOSAICS model. 
As the formation of massive clusters requires high natal gas densities and pressures \citep[see][]{Elmegreen_and_Efremov_97, Kruijssen_15, P18}, UV-bright young GCs therefore typically only occur in galaxies with currently high SFRs, which are therefore also very UV-bright (Fig.~\ref{fig:formtime}).

\subsection{The role of young GCs in reionization}
\label{sec:reionization}

A number of works have suggested that young GCs may play a significant role in reionization \citep[e.g.][]{Ricotti_02, Griffen_et_al_13, Katz_and_Ricotti_13, Katz_and_Ricotti_14, Boylan-Kolchin_17, Boylan-Kolchin_18}.
Based on measurements of the Thomson optical depth, the average redshift of reionization has been found to be between $z=7.8$ and $8.8$ \citep{Planck_2016_paperXLVII} and was inferred to be fully completed completed by $z=5.5$ \citep{Becker_et_al_15, McGreer_Mesinger_and_DOdorico_15}.
The interpretation that GCs significantly contributed to reionization therefore crucially depends on the assumptions that (metal-poor) GC formation occurred at redshifts $z\gtrsim6$ and that GCs dominate the ionizing radiation at these redshifts.

These assumptions can be tested with the E-MOSAICS simulations, under the ansatz that GCs and observed young star clusters (today) have the same formation mechanism.
In the simulations of Milky Way-mass galaxies, the majority of GCs (even metal-poor ones) form after reionization was completed \citep{Reina-Campos_et_al_19} and therefore most GCs do not make any contribution to reionization.
For present-day galaxies more massive than the Milky Way, whose star formation histories are shifted to earlier epochs \citep{Qu_et_al_17}, a larger fraction of GCs may form prior to the epoch of reionization (which remains to be tested in the E-MOSAICS model). 
However, the majority of GCs in the Universe are expected to form in the progenitors of $\sim L^\ast$ (approximately Milky Way-mass) galaxies \citep{Harris_16}, and therefore the simulations should be representative of the typical formation histories of GC populations.
For those clusters that do form prior to the epoch of reionization, the typical CFEs of a few tens of per cent at high redshifts \citep[fig. 6 in][]{P18} also means that GCs do not dominate the total UV flux in a galaxy (with massive clusters contributing an even smaller fraction due to the low-mass power law of the initial cluster mass function).
Indeed, in this work, we find that the star clusters typically contribute less than 10 per cent of the UV flux in a galaxy, even at redshifts $z>6$. Therefore, assuming similar escape fractions of ionizing photons for both field stars and star clusters, our work suggests GCs should only make a sub-dominant contribution to reionization.

However, for the GCs that form during the epoch of reionization, their contribution crucially depends on the escape fraction of ionizing photons relative to that of the field stars.
This in turn depends on the structure of the interstellar medium and whether star clusters have a significantly higher escape fraction than for the field stars.
Young clusters locally are observed to be gas free on short timescales \citep[few Myr,][]{Bastian_Hollyhead_and_Cabrera-Ziri_14, Hollyhead_et_al_15, Kruijssen_et_al_19c} which might imply high escape fractions.
However, if the star-forming complexes within which the GCs form also become gas free on similar timescales, then the field stars and GCs will have a similar escape fraction within a given galaxy \citep[which is suggested if the UV-bright objects at high redshifts are indeed cluster complexes,][]{Bouwens_et_al_17c}.
Answering whether GCs significantly contributed to reionization therefore requires the self-consistent treatment of the interstellar and intragalactic medium, star and GC formation, ionization through radiative transfer and other stellar feedback processes like stellar winds and supernovae, and is well beyond the scope of this work.

\section{Conclusions}
\label{sec:conclusions}

High-redshift observations of young GCs potentially offer a powerful way to test the formation theories of GCs, since they do not depend on the uncertain relation between present-day age measurements and the initial properties of the GC population.
In this work, we analyse the rest-frame UV and optical properties of the GC populations in the E-MOSAICS simulations of Milky Way-mass haloes.
We find that the most massive clusters are rarely the brightest clusters in the UV, due to the rapid fading of the stellar populations. 
The typical brightest clusters in the progenitors of Milky Way-mass galaxies vary with redshift from $\muv \approx -11$ at $z=6$ and peak at $\muv \approx -14$ at $z=1$-$2$.
This evolution is driven by the change of the cluster formation rate with redshift, peaking at a similar time as the clumpy fraction of galaxies \citep{Guo_et_al_15, Shibuya_et_al_16}.
The brightest clusters of all populations are consistent with the objects found by \citet{Vanzella_et_al_17a}, although contamination by cluster complexes may affect the observations.

Using observations of young clusters and clusters complexes in nearby galaxies, we calculated the effect of cluster complexes on UV observations of unresolved clusters. 
We find that at apertures of $50$ and $100 \pc$ the complexes are $\sim0.8$ and $\sim1.1$ magnitudes brighter than the clusters within them, respectively (though with significant scatter between individual regions). 
This suggests that high-redshift observations do not retrieve individual clusters, but cluster complexes \citep[see also][]{Bouwens_et_al_17c}.

By calculating the UV luminosities of the GC host galaxies, we determine the fraction of UV flux that young GCs contribute to galaxies at different redshifts. 
We find that clusters typically contribute $<10$ per cent of the UV flux in a galaxy at all redshifts, with a maximum of $\sim70$ per cent. At $z>2$, only $0.2$ per cent of galaxies have a UV flux contribution from clusters larger than $0.5$.
The single brightest cluster in a galaxy typically contributes $<5$ per cent of the total UV flux, due to the low chance of observing massive clusters at very young ages ($<10 \Myr$).
Overall, the maximum instantaneous value of the UV fraction contributed by GCs is set by the CFE, while cluster formation is biased to periods of high SFRs when the galaxy is UV-bright, meaning GCs are never expected to dominate the UV flux in a galaxy.

Due to their compact sizes, GCs could be significantly easier to detect than the more extended emission of the galaxy, despite not dominating the total UV flux.
Under reasonable assumptions of the sizes of high-redshift galaxies (based on observations), we find that GCs do not generally have significantly higher surface brightnesses than their host galaxies, because GCs form when galaxies are UV-bright and massive GCs are unlikely to be observed at extremely young ages.
At $z>6$ and $30\pc$ resolution, GCs are typically only $\approx 1 \Mag \Arcsec^{-2}$ brighter than the peak surface brightness of the host galaxy.
Due to the increasing sizes of the galaxies with cosmic time, young clusters are typically easier to detect at low redshifts at a given resolution of observation.
However, these results present a lower limit, because it is not possible to account for the contribution of cluster complexes to cluster brightnesses.

Finally, we discuss the potential role of GCs in the reionization of the Universe, finding that due to the formation times of most GCs after reionization and the low contribution of GCs to the total UV flux, GCs should not make a major contribution to reionization.
However, this conclusion does depend on the relative escape fraction of ionizing photons between GCs and field stars in a galaxy. This value is unconstrained and therefore presents an important avenue for future work.

\section*{Acknowledgments}

We thank the anonymous referee for helpful and constructive comments which improved the paper.
JP and NB gratefully acknowledge financial support from the European Research Council (ERC-CoG-646928, Multi-Pop).  NB and RAC are Royal Society University Research Fellows. JMDK gratefully acknowledges funding from the German Research Foundation (DFG) in the form of an Emmy Noether Research Group (KR4801/1-1). JMDK and MRC gratefully acknowledge funding from the European Research Council (ERC-StG-714907, MUSTANG). MRC is supported by a Fellowship from the International Max Planck Research School for Astronomy and Cosmic Physics at the University of Heidelberg (IMPRS-HD). This work used the DiRAC Data Centric system at Durham University, operated by the Institute for Computational Cosmology on behalf of the STFC DiRAC HPC Facility (www.dirac.ac.uk). This equipment was funded by BIS National E-infrastructure capital grant ST/K00042X/1, STFC capital grants ST/H008519/1 and ST/K00087X/1, STFC DiRAC Operations grant ST/K003267/1 and Durham University. DiRAC is part of the National E-Infrastructure. This study also made use of high performance computing facilities at Liverpool John Moores University, partly funded by the Royal Society and LJMUs Faculty of Engineering and Technology.


\bibliographystyle{mnras}
\bibliography{emosaics}

\bsp
\label{lastpage}
\end{document}